\documentclass[aps,11pt,nofootinbib,preprintnumbers,showpacs,superscriptaddress]{revtex4}

 \usepackage{here}

\usepackage[dvipdfmx]{graphicx}
\usepackage{amssymb,amsfonts,amsmath,bm,color,multirow,cases,empheq,hyperref}
\usepackage{mathrsfs}

\usepackage{enumerate}
\usepackage{ulem}
\usepackage{afterpage}

\hypersetup{%
 setpagesize=false,
 bookmarksnumbered=true,%
 bookmarksopen=true,%
 colorlinks=true,%
 linkcolor=blue,%
 citecolor=red}

\begin{document}




\title{
Lower bound for angular momenta of microstate geometries in five dimensions
}

\author{Shinya Tomizawa}
\email{tomizawa@toyota-ti.ac.jp}
\affiliation{Mathematical Physics Laboratory, Toyota Technological Institute, Nagoya 468-8511, Japan}

\preprint{TTI-MATHPHYS-5}




\begin{abstract} 
 We study the  Bogomol'nyi-Prasad-Sommerfield (BPS) solutions of the asymptotically flat, stationary microstate geometries with bi-axisymmetry and reflection symmetry in the five-dimensional ungauged minimal supergravity.      
 We show that the angular momenta of the microstate geometry with a small number of centers (at least, five centers) have lower bounds, which are slightly smaller than those of the maximally spinning BMPV black hole. 
Therefore, there exists a certain narrow parameter region such that the microstate geometry with a small number of the centers admits the same angular momenta as the BMPV black hole.  
Moreover, we investigate the dependence of the topological structure of the evanescent ergosurfaces on the magnetic fluxes through the $2$-circles between two centers. 

\end{abstract}

\pacs{04.50.+h  04.70.Bw}
\date{\today}
\maketitle



\section{Introduction}

The microstate geometries~
\cite{Lunin:2001jy,Maldacena:2000dr,Balasubramanian:2000rt,Lunin:2002iz,Lunin:2004uu,Giusto:2004id,Giusto:2004ip,Giusto:2004kj,Bena:2005va,Gibbons:2013tqa,Saxena:2005uk,Skenderis:2008qn,Balasubramanian:2008da,Chowdhury:2010ct} are smooth horizonless solutions in the bosonic sector of supergravity which have the same asymptotic structure as a given black hole or a black ring. 
So far, these solutions have been constructed and thought of as one of ways to resolve  the problem of black hole information loss. 
This idea to describe black hole microstates by horizonless geometries originated from the works on fuzzballs of Mathur~\cite{Mathur:2005zp,Mathur:2005ai,Mathur:2008nj}.  
The existence of such solutions itself should be surprising because of the earlier results \cite{Einstein1941,Einstein1943,Breitenlohner:1987dg,Cater1986} on ``No-Go" which exclude completely smooth soliton solutions which are regular in four spacetime dimensions. 
In five dimensional supergravity, the conclusion of the no-go theorem can be evaded because the spacetime admits the spatial cross sections with non-trivial second homology and the Chern-Simons interactions.

\medskip
Therefore, despite the absence of horizons, the microstate geometries should closely approximate the geometries of black holes and need to describe all phenomenon which could occur in black hole spacetimes, such as the gravitational lens effect and gravitational wave radiation.  
However, the analysis is not still sufficient to say that such microstate geometries well describes black hole physics. From this point of view, it is an important issue to probe what extent asymptotically flat microstate geometries  possess the classical features of stationary black holes with the same asymptotic structure. 
There are many ways to probe physical aspects of such microstate geometries. 
A natural and simple way is to investigate whether these spacetimes can carry the same asymptotic charges, the mass and angular momenta, as rotating black holes~\cite{Tangherlini:1963bw,Myers:1986un,Breckenridge:1996is}, rotating black rings~\cite{Emparan:2001wn,Pomeransky:2006bd,Elvang:2004rt,Elvang:2004ds} and rotating black lenses~\cite{Kunduri:2014kja,Tomizawa:2016kjh,Breunholder:2017ubu,Kunduri:2016xbo,Tomizawa:2019yzb} in the same theory.  
If not so,  such spacetimes cannot be regarded as the description of these black objects. 
For instance, as proved mathematically in~\cite{Kunduri:2017htl}, there are no asymptotically static microstate geometries in higher dimensional Einstein-Maxwell theory, which implies that  any static black hole cannot be described by the soliton solutions. 
In particular, it is well known that there exist the microstate geometries corresponding to maximally-spinning black holes and maximally-spinning black rings that have zero horizon area, which are referred to “zero-entropy microstate geometries"~\cite{Bena:2006is}. 
Moreover, using the merge of such zero-entropy microstate geometries, Refs.~\cite{Bena:2006kb,Bena:2007qc} constructed the first microstate geometries with the same charges as black holes and  black rings which have nonzero horizon area. 
In general,  it is, however, not known how to construct the microstate geometries that correspond to black holes and black rings with non-zero horizon area without  introducing  the merger of zero-entropy microstate geometries.

\medskip
The main purpose of this paper is to investigate whether there exist the microstate geometries in five dimension having the same asymptotic charges (mass and  angular momenta) as the black hole, without using zero-entropy microstate geometries and by merely imposing a simple symmetry. 
In this paper, based on the work developed by Gauntlett {\it et al.}~\cite{Gauntlett:2002nw}  in the framework of the five-dimensional minimal ungauged supergravity, we consider  asymptotically flat, stationary and bi-axisymmetric BPS microstate geometries with $n$ centers  on the $z$-axis of the Gibbons-Hawking space.
In addition, we impose reflection symmetry, which means the invariance under the transformation $z\to - z$,  on the solution since such an assumption dramatically simplifies the constraint equations for the parameters included in the solutions, a so-called “bubble equations", and this enables us to solve the constraint equations for the parameters. 
It can be shown that under the symmetry assumptions, the geometry has equal angular momenta. 
It is of physical interest to compare the mass and angular momenta of the Breckenridge-Myers-Peet-Vafa (BMPV)  solution~\cite{Breckenridge:1996is} since it describes a spinning black hole with equal angular momenta in the same theory. 
We will show that asymptotically flat, stationary, bi-axisymmetric and reflection-symmetric microstate geometries (at least, for five centers) can have the same mass and angular momenta as  the BMPV black hole.

\medskip
The rest of the paper is  organized as follows: 
In the following Sec.~\ref{sec:msg}, we review the BPS solutions of the microstate geometries in the five-dimensional minimal supergravity. 
In Sec.~\ref{sec:analysis}, we compute the mass, angular momenta and magnetic fluxes through the bubbles, and show the existence  of evanescent ergosurfaces.  
In Sec.~\ref{sec:reflection}, imposing reflection symmetry, we simplify the solution and the bubble equations and thereafter show numerically that the microstate geometries have the same angular momentum  as the BMPV black hole.
In Sec.~\ref{sec:summary}, we summarize our results and discuss possible generalizations of our analysis.

\section{Microstate geometry}\label{sec:msg}

\subsection{Solutions}

First, 
we begin with supersymmetric solutions in the five-dimensional minimal ungauged supergravity~\cite{Gauntlett:2002nw}, whose bosonic Lagrangian consists of the Einstein-Maxwell theory with a Chern-Simons term.
In this theory, the metric and the gauge potential of Maxwell field for the supersymmetric solutions take the form:
\begin{eqnarray}
\label{metric}
ds^2&=&-f^2(dt+\omega)^2+f^{-1}ds_{M}^2,\label{eq:solution_metric}\\
A&=&\frac{\sqrt 3}{2} \left[f(d t+\omega)-\frac KH(d \psi+\chi)-\xi \right]. \label{eq:solution_1form}
\end{eqnarray}
Here, the four-dimensional metric $ds^2_M$ is the metric of an arbitrary hyper-K\"{a}hler space, where we use the Gibbons-Hawking space metric~\cite{Gibbons:1979zt} which is written as
\begin{eqnarray}
ds^2_M&=&H^{-1}(d\psi+\chi)^2+Hds^2_{{\mathbb E}^3}, \label{eq:GH}\\
ds^2_{{\mathbb E}^3}&=&dx^2+dy^2+dz^2,\\
H&=&\sum_{i=1}^n\frac{h_i}{r_i}, \label{Hdef}
\end{eqnarray}
with
\begin{eqnarray}
r_i:&=&|{\bm r}-{\bm r_i}|=\sqrt{(x-x_i)^2+(y-y_i)^2+(z-z_i)^2},\\
{\bm r}:&=&(x,y,z),\\
{\bm r}_i:&=&(x_i,y_i,z_i),
\end{eqnarray}
The function $H$ in Eq.(\ref{Hdef}) is a harmonic function with $n$ point sources ($n$ centers) on three-dimensional Euclid space ${\mathbb E}^3$, and the 1-form $\chi$ on ${\mathbb E}^3$ is determined by
\begin{eqnarray}
*d\chi=dH, 
\end{eqnarray}
where the Hodge dual $*$ is associated with ${\mathbb E}^3$. 
$\chi$ can be written as
\begin{eqnarray}
\chi&=&\sum_{i=1}^nh_i\tilde \omega_i,
\end{eqnarray}
where the 1-form $\tilde\omega_i$ on ${\mathbb E}^3$, which is defined by
\begin{eqnarray}
*d\tilde \omega_i=d(1/r_i),
\end{eqnarray}
can be written as
\begin{eqnarray}
\tilde \omega_i=\frac{z-z_i}{r_i}
\frac{(x-x_i)dy-(y-y_i)dx}{(x-x_i)^2+(y-y_i)^2}.
\end{eqnarray}
The vectors $\partial/\partial t$ and $\partial/\partial \psi$ are commuting Killing vector fields. 
The Gibbons-Hawking  metric~(\ref{eq:GH}) is preserved under the scaling transformation $H\to \lambda^2 H$, $\chi \to \lambda^2 \chi$, $\psi\to \lambda\psi $ and $x^i\to \lambda^{-1} x_i$, which enables us to fix the period of the coordinate $\psi$ as $0\le\psi <4\pi$. 
These Gibbons-Hawking spaces are nontrivial $U(1)$ fibration over a flat space ${\mathbb E}^3$ and the unique class of four-dimensional hyper-K\"{a}hler metric with tri-holomorphic isometry.

The function $f^{-1}$ and the 1-forms $(\omega,\xi)$ are given by
\begin{eqnarray}
f^{-1}&=&H^{-1}K^2+L,\\
\omega&=&\omega_\psi(d\psi+\chi)+\hat \omega,\\
\omega_\psi&=&H^{-2}K^3+\frac{3}{2} H^{-1}KL+M, 
\end{eqnarray}
where the functions $K$, $L$ and $M$ are harmonic functions on ${\mathbb E}^3$, 
\begin{eqnarray}
K&=&\sum_{i=1}^n\frac{k_i}{r_i},\\
L&=&l_0+\sum_{i=1}^n\frac{l_i}{r_i}, \\
M&=&m_0+\sum_{i=1}^n\frac{m_i}{r_i},
\end{eqnarray}
The 1-forms $\hat \omega$ are $\xi$ are determined by
\begin{eqnarray}
*d\hat\omega&=&HdM-MdH+\frac{3}{2}(KdL-LdK), \label{eq:defomega}\\
*d\xi&=&-dK,
\end{eqnarray}
and take the forms
\begin{eqnarray}
\hat \omega&=&\sum_{i,j=1(i\not=j)}^n\left(h_i m_j+\frac{3}{2}k_il_j\right)\hat\omega_{ij}-\sum_{i=1}^n\left( m_0h_i+\frac{3}{2}l_0k_i \right)\tilde\omega_i,\\
\xi&=&-\sum_{i=1}^n k_i \tilde\omega_i,
\end{eqnarray}
where the 1-form $\hat\omega_{ij}\ (i\not=j)$ on ${\mathbb E}^3$, which is determined by
\begin{eqnarray}
*d\hat \omega_{ij}=(1/r_i)d(1/r_j)-(1/r_j)d(1/r_i),
\end{eqnarray}
can be written as
\begin{eqnarray}
\hat\omega_{ij}=-\frac{({\bm r}-{\bm r}_i)\cdot ({\bm r}-{\bm r}_j)}{r_ir_j}\frac{\left[ ({\bm r}_i-{\bm r}_j)\times ({\bm r}-\frac{{\bm r_i}+{\bm r_j}}{2})\right]_kdx^k}{\left|({\bm r}_i-{\bm r}_j)\times({\bm r}-\frac{{\bm r_i}+{\bm r_j}}{2})\right|^2}.
\end{eqnarray}

\medskip
In this paper, we set ${\bm r}_i=(0,0,z_i)$ $(i=1,\ldots,n)$, by which $x\partial/\partial y-y\partial/\partial x$ becomes another $U(1)$ Killing vector field, and assume $z_i<z_j$ for $i<j$  $(i ,j=1,\ldots,n)$ without loss of generality. 
In terms of standard spherical coordinates $(r,\theta,\phi)$ such that $(x,y,z)=(r\sin\theta\cos\phi,r\sin\theta\sin\phi,r\cos\theta)$, the 1-forms  $\tilde\omega_i$ and $\hat\omega_{ij}$ are simplified as
\begin{eqnarray}
\tilde \omega_i&=&\frac{r\cos\theta -z_i}{r_i}d\phi,\\
\hat\omega_{ij}&=&\frac{r^2-(z_i+z_j)r\cos\theta+z_iz_j}{z_{ji}r_ir_j}d\phi,\quad  z_{ji}:=z_j-z_i.
\end{eqnarray}
and so the  1-form $\hat\omega$ can be written as
\begin{eqnarray}
\hat \omega&=&\biggl[\sum_{i,j=1}^n\left(h_i m_j+\frac{3}{2}k_il_j\right)\frac{r^2-(z_i+z_j)r\cos\theta+z_iz_j}{z_{ji}r_ir_j}\notag\\
&&-\sum_{i=1}^n\left( m_0h_i+\frac{3}{2}l_0k_i \right)\frac{r\cos\theta -z_i}{r_i}+c\biggr]d\phi,
\end{eqnarray}
where we have added the integration constant $c$ since $\hat \omega$ is determined by only the derivatives in Eq.~(\ref{eq:defomega}).

\medskip
\subsection{Boundary conditions}



As the detail is reviewed in~\cite{Gibbons:2013tqa,Warner:2019jll}, 
in order that the supersymmetric solution describes the BPS microstate geometry solution of physical interest, 
we must  impose suitable boundary conditions 
 (i) at infinity, 
(ii) at the Gibbon-Hawking centers ${\bm r}={\bm r}_i\ (i=1,...,n)$ and 
(iii) on the $z$-axis $x=y=0$ of ${\mathbb E}^3$ in the Gibbons-Hawking space.  
More precisely, we consider the following boundary conditions: 

\begin{description}
\item[ (i)] at infinity $r\to\infty$, the spacetime is asymptotically Minkowski spacetime. 
\item[(ii)] at  the $n$ centers ${\bm r}={\bm r}_i\ (i=1,...,n)$ such that each harmonic function diverges,  the spacetime is regular, and behaves as the coordinate singularities like the origin of the Minkowski spacetime. 
The spacetime admits no causal pathology such as closed timelike curve (CTCs) around these points.  
\item[(iii)] on the $z$-axis  $I=\{(x,y,z)\ |\ x=y=0\}$ of ${\mathbb E}^3$ in the Gibbons-Hawking space,  there appear no Dirac-Misner strings,  no orbifold singularities and no conical singularities. 
\end{description}

\subsubsection{Infinity}
The asymptotic flatness demands that at infinity $r\to\infty$, the functions ($f^{-1}$, $H$), the 1-forms ($\chi$, $\omega$) behave as, respectively, 
\begin{eqnarray}
f^{-1} &\simeq& 1,\label{eq:asymptotic_fi}\\
\quad H&\simeq& \frac{1}{r}, \label{eq:asymptotic_H}\\
 \omega &\simeq& 0,\label{eq:asymptotic_omega}\\
  \chi &\simeq& \pm\cos\theta d\phi,  \label{eq:asymptotic_chi}
\end{eqnarray}
which ensure that in terms of the radial coordinate $\rho=2\sqrt{r}$, and at  $r\to\infty$ ($\rho \to \infty$)
the metric is indeed approximated as 
\begin{eqnarray}
ds^2&\simeq& -dt^2+d\rho^2+\frac{\rho^2}{4}\left[(d\psi+\cos\theta d\phi)^2+ d\theta^2+\sin^2\theta d\phi^2 \right].\label{eq:asymptotic_metric}
\end{eqnarray}
This is  the metric of five-dimensional Minkowski spacetime where the metric on $S^3$ is written in terms of Euler angles ($\psi, \phi, \theta)$, whose ranges must be $0\le \theta \le \pi$, $0\le \phi <2\pi $ and $0\le \psi <4\pi$ with the identification $\phi\sim \phi+2\pi$ and $\psi\sim \psi+4\pi$.

\medskip
At infinity $r \to \infty$, the metric functions $f$ and $H$ behave, respectively,  as 
\begin{eqnarray}
f^{-1}&\simeq &l_0+\left[\left(\sum_{i}k_i\right)^2+\sum_{i}l_i\right]\left(\sum_i h_i\right)^{-1}  r^{-1}, \label{eq:infinity-f} \\
 H&\simeq &\left(\sum_i h_i\right)r^{-1} \label{eq:infinity-H}. 
 \end{eqnarray}
Since for $r\to \infty$, the metric function  $\omega_\psi$  and the $1$-forms ($\tilde \omega_i$, $\hat \omega_{ij}$) behave as, respectively, 
\begin{eqnarray}
\omega_\psi &\simeq& m_0+\frac{3}{2}l_0\left(\sum_i h_i\right)^{-1}\sum_{i} k_i, \\
\tilde \omega_i  &\simeq& \cos\theta d\phi,\\
\hat\omega _{ij} &\simeq& \frac{d\phi}{z_{ji}},
\end{eqnarray}
 the 1-forms $\chi$ and $\omega$ are approximated by
\begin{eqnarray}
\chi&=& \sum_{i} h_i \hat\omega_i\simeq \sum_{i} h_i\cos\theta d\phi, \label{eq:infinity-chi}\\
\omega
                    &\simeq & \left(m_0+\frac{3}{2}l_0\sum_{i}k_i\right)\left(d\psi +\cos\theta d\phi \right)-\sum_{i}\left(m_0h_i+\frac{3}{2}l_0k_i\right)\cos\theta d\phi \notag \\
                    &&+\left(\sum_{i,j(i\not =j)}\frac{h_im_j
                    +\frac{3}{2}k_il_j}{z_{ji}} +c\right)d\phi. \label{eq:infinity-omega}
\end{eqnarray}
Thus, in comparison with Eqs.~(\ref{eq:asymptotic_fi})-(\ref{eq:asymptotic_chi}) and Eqs.~(\ref{eq:infinity-f}), (\ref{eq:infinity-H}),(\ref{eq:infinity-chi}), (\ref{eq:infinity-omega}), the parameters must  satisfy the following constraints
\begin{eqnarray}
l_0&=&1,\label{eq:l0}\\
\sum_{i=1}^n h_i&=&1, \label{eq:sumhi}\\
m_0&=&-\frac{3}{2}l_0\sum_{i=1}^nk_i,\label{eq:m0}\\
c&=&-\sum_{i,j=1(i\not =j)}^n\frac{h_im_j+\frac{3}{2}k_il_j}{z_{ji}}.\label{eq:c}
\end{eqnarray}

\subsubsection{Gibbons-Hawking centers}
The metric obviously has divergence at the points ${\bm r}={\bm r}_i$ ($n=1,...,n$) on the Gibbons-Hawking space. 
We hence impose the boundary conditions at the points ${\bm r}={\bm r}_i$ ($n=1,...,n$) so that these become regular points like the origin of Minkowski spacetime: 
\begin{eqnarray}
ds^2\simeq  -dt'{^2}+\left[d\rho^2+\frac{\rho^2}{4}\left\{(d\psi'\pm\cos\theta d\phi')^2+d\theta^2+\sin^2\theta d\phi'{}^2\right\}\right] \,. \label{eq:metric@center1}
\end{eqnarray}

Let us choose the coordinates $(x,y,z)$ on ${\mathbb E}^3$ of the Gibbons-Hawking space so that the $i$th point ${\bm r}={\bm r}_i$ is an origin of ${\mathbb E}^3$. 
Near the origin ${\bm r}=0$, the four harmonic functions $H$, $K$, $L$ and $M$ behave as, respectively,   
\begin{eqnarray}
H&\simeq&\frac{h_i}{r}+\sum_{j(\not=i)}\frac{h_j}{|z_{ji}|},\quad K\simeq\frac{k_i}{r}+\sum_{j(\not=i)}\frac{k_j}{|z_{ji}|},\quad \\
L&\simeq&\frac{l_i}{r}+1+\sum_{j(\not=i)}\frac{l_j}{|z_{ji}|},\quad  M\simeq\frac{m_i}{r}+m_0+\sum_{j(\not=i)}\frac{m_j}{|z_{ji}|}, 
\end{eqnarray}
which lead to
\begin{eqnarray}
f^{-1}&\simeq& \frac{k_i^2h_i^{-1}+l_i}{r}+c_{1(i)},\\
\omega_\psi&\simeq& \frac{k_i^3h_i^{-2}+\frac{3}{2}k_il_ih_i^{-1}+m_i}{r}+c_{2(i)},
\end{eqnarray}
where the constants $c_{1(i)}$ and $c_{2(i)}$ are defined by
\begin{eqnarray}
h_ic_{1(i)} &:=&h_il_0+\sum_{j=1(j\not=i)}^n \frac{2h_i^2k_ik_j-h_ik_i^2h_j+h_i^3l_j}{|z_{ij}|h_i^2}\notag \label{eq:c1}\\
&=&h_i+\sum_{j=1(j\not=i)}^n \frac{2k_ik_j-h_ik_i^2h_j-h_ih_jk_j^2}{|z_{ij}|} ,\label{eq:c1b}\\
h_ic_{2(i)}&:=&h_im_0+\frac{3}{2}k_il_0+\sum_{j=1(j\not=i)}^n\frac{-2k_i^3h_j+3h_ik_i^2k_j+3h_i^2k_il_j+2h_i^3m_j}{2|z_{ij}|h_i^2}\notag\label{eq:c2}\\
&=&h_im_0+\frac{3}{2}k_i+\sum_{j=1(j\not=i)}^n\frac{- 2m_ih_j-3l_ik_j+3k_il_j+2h_im_j}{2|z_{ij}|},\label{eq:c2b}
\end{eqnarray}
where we have used $h_i^2=1$ ($h_i=\pm 1$ will be imposed below. See Eq.~(\ref{eq:hi}))  in the second equalities of Eqs.(\ref{eq:c1b}) and (\ref{eq:c2b}). 
The 1-forms $\tilde \omega_j$ and $\hat\omega_{kj}$ are approximated by 
\begin{eqnarray}
&&\tilde \omega_i\simeq \cos\theta d\phi,\qquad 
\tilde \omega_j\simeq -\frac{z_{ji}}{|z_{ji}|} d\phi\ (j\not=i),\\
&&\hat \omega_{ij}\simeq -\frac{\cos\theta}{|z_{ji}|} d\phi\ (i\not= j) ,\qquad \hat \omega_{kj}\simeq \frac{z_{ji}z_{ki}}{|z_{ji}z_{ki}|z_{jk}} d\phi\ (k\not =j, k,j\not=i), 
\end{eqnarray}
and hence, 1-forms $\chi$ and $\hat\omega$ behave as
\begin{eqnarray}
\chi&\simeq& \left(h_i\cos\theta+\chi_{0(i)}\right)d\phi\,, \qquad 
\hat\omega\simeq(\hat \omega_{1(i)}\cos\theta+\hat\omega_{0(i)})d\phi \,,
\end{eqnarray}
where 
\begin{eqnarray}
\chi_{0(i)} & := &- \sum_{j(\not=i)}\frac{h_jz_{ji}}{|z_{ji}|} \,,\\
\hat \omega_{0(i)}& := &  \sum_{k,j(\not=i,k\not=j)}\left(h_km_j+\frac{3}{2}k_kl_j\right)\frac{z_{ji}z_{ki}}{|z_{ji}z_{ki}|z_{jk}}+\sum_{j(\not= i)}\left(m_0h_j+\frac{3}{2}k_j\right)\frac{z_{ji}}{|z_{ji}|}+c\,, \\
\hat \omega_{1(i)}& := & -\sum_{j(\not=i)}\left(h_im_j-h_jm_i+\frac{3}{2}(k_il_j-k_jl_i)\right)\frac{1}{|z_{ji}|}-\left(m_0h_i+\frac{3}{2}k_i\right)\,.
\end{eqnarray}
One therefore obtains the asymptotic behavior of the metric around the $i$th point  as
{\small
\begin{eqnarray}
ds^2&\simeq& -\left( \frac{k_i^2h_i^{-1}+l_i}{r}+c_{1(i)}\right)^{-2}\biggl[dt+\left(\frac{k_i^3h_i^{-2}+\frac{3}{2}k_il_ih_i^{-1}+m_i}{r}+c_{2(i)}\right)\left\{d\psi+(h_i\cos\theta+\chi_{0(i)})d\phi \right\} \nonumber\\
&&+(\hat \omega_{1(i)}\cos\theta+\hat\omega_{0(i)})d\phi \biggr]^2+\left( \frac{k_i^2h_i^{-1}+l_i}{r}+c_{1(i)}\right)\frac{r}{h_i}\nonumber \\
&&\times\left[\left\{d\psi+(h_i\cos\theta+\chi_{0(i)})d\phi \right\}^2+h_i^2\left(\frac{dr^2}{r^2}+d\theta^2+\sin^2\theta d\phi^2\right)\right].
\end{eqnarray}
}
To remove the divergence of the metric, it is sufficient to impose the following conditions on the parameters $(k_i,l_i,m_i)\ (i=1,...,n)$:
\begin{eqnarray}
&&k_i^2+h_il_i=0, \\
&& k_i^3h_i^{-2}+\frac{3}{2}k_il_ih_i^{-1}+m_i=0,
\end{eqnarray}
which are equivalent to the condition for the parameters $(l_i,m_i)$,
\begin{eqnarray}
l_i&=&-\frac{k_i^2}{h_i},\label{eq:condition1}\\
m_i&=&\frac{k_i^3}{2h_i^2}\label{eq:condition2},
\end{eqnarray}
and these yield the equation
\begin{eqnarray}
h_ic_{2(i)}=-\hat\omega_{1(i)}.
\end{eqnarray}
Introducing the new coordinates $(\rho,\psi', \phi')$ by
\begin{eqnarray}
\rho=2\sqrt{h_i^{-1}c_{1(i)}r}, \qquad  \psi'=\psi+\chi_{0(i)}\phi, \qquad \phi'=\phi \,, 
\end{eqnarray}
we can write the metric near  ${\bm r}={\bm r}_i$ as
\begin{eqnarray}
ds^2\simeq -c_{1(i)}^{-2} d[t+c_{2(i)}\psi'+\hat \omega_{0(i)}\phi']^2+\left[d\rho^2+\frac{\rho^2}{4}\left\{(d\psi'+h_i\cos\theta d\phi')^2+d\theta^2+\sin^2\theta d\phi'{}^2\right\}\right] \,. \label{eq:metric@center2}
\label{nutmetric}
\end{eqnarray}
Comparing the $(\phi',\psi')$-part of the above metric~(\ref{eq:metric@center2}) with the boundary condition~(\ref{eq:metric@center1}), we must impose for each $h_i\ (i=1,\ldots,n)$
\begin{eqnarray}
h_i=\pm 1. \label{eq:hi}
\end{eqnarray}

To ensure the five-dimensional metric with Lorentzian signature, the following inequities must be satisfied
\begin{eqnarray}
h_i^{-1}c_{1(i)}=h_i+\sum_{j=1(j\not=i)}^n\frac{2k_ik_j+l_ih_j+h_il_j}{|z_{ij}|}
                     >0 \ (i=1,\ldots,n) .\label{eq:c1ineq}
\end{eqnarray}
The above metric~(\ref{nutmetric}) is locally isometric to the flat metric, but 
CTCs necessarily appear near $\rho \simeq 0$ because the Killing vector $\partial/\partial \psi'=\partial/\partial \psi$ becomes timelike. 
To avoid the existence of CTCs around ${\bm r}_i\ (i=1,\ldots,n)$,  $c_{2(i)}=0$ and $\omega_{0(i)}=0$ must be simultaneously satisfied at ${\bm r}={\bm r}_i$ ($i=1,...,n$) but it is sufficient to impose only $c_{2(i)}=0$,  which can be written as
\begin{eqnarray}
0&=&h_ic_{2(i)}\notag\\
 &=&h_im_0+\frac{3}{2}k_i+\sum_{j=1(j\not=i)}^n\frac{h_im_j-m_ih_j-\frac{3}{2}(l_ik_j-k_il_j)}{|z_{ij}|}\notag\\
&=&h_im_0+\frac{3}{2}k_i+\sum_{j=1(j\not=i)}^n\frac{(h_ik_j-h_jk_i)^3}{2|z_{ij}|}.
\label{eq:condition3b}
\end{eqnarray}
These equations are so-called ``bubble equations'' in Refs.~\cite{Bena:2005va,Bena:2007kg}, 
which physically means the balance between the gravitational attraction and the repulsion by the magnetic fluxes over the 2-cycles. 
Moreover, let us note that $\hat\omega_{0(i)}=0$ automatically hold for all $i=1,...,n$,  if we impose (\ref{eq:condition3b}) since from Eqs.~(\ref{eq:condition1}) and (\ref{eq:condition2}), 
$\hat\omega_{0(i)}$ can be shown to vanish, 
\begin{eqnarray}
\hat\omega_{0(i)}&=&\sum_{k,j(k,j\not=i,k\not=j)}\left(h_km_j+\frac{3}{2}k_kl_j\right)\frac{z_{ji}z_{ki}}{|z_{ji}z_{ki}|z_{jk}}+\sum_{j(\not=i)}\left( m_0h_j+\frac{3}{2}k_j\right)\frac{z_{ji}}{|z_{ji}|}\nonumber \notag\\
&&-\sum_{k,j(j\not=k)}\frac{h_km_j+\frac{3}{2}k_kl_j}{z_{jk}} \label{eq:omega0-1}\notag\\
&=&\sum_{k,j(k,j\not=i,k\not=j)}\frac{(h_kk_j-h_jk_k)^3}{4z_{jk}} \frac{z_{ji}z_{ki}}{|z_{ji}z_{ki}|}
+\sum_{k,j(j\not=i,k\not=j)}\frac{(h_kk_j-h_jk_k)^3}{2|z_{jk}|} \frac{z_{ji}}{|z_{ji}|}\notag\\
&&-\sum_{k,j(k\not=j)}\frac{(h_kk_j-h_jk_k)^3}{4z_{jk}}\notag\\
&=&0,
\label{eq:omega}
\end{eqnarray}
where we have used Eq.~(\ref{eq:condition3b}) for the 2nd term in the right-hand side of the first equality, and the last equality can be shown by long but simple computations.

Furthermore, the $n$ bubble equations $c_{2(i)}=0\ (i=1,\ldots,n)$ are not independent because the summation of $h_ic_{2(i)}\ (i=1,\ldots,n)$  automatically  vanishes, regardless of the bubble equations, as
  \begin{eqnarray}
\sum_{i=1}^nh_ic_{2(i)}&=&\sum_{i=1}^nh_im_0+\frac{3}{2}\sum_{i=1}^nk_i \notag\\
                                  &&+\sum_{i=1}^n\sum_{j=1(j\not=i)}^n\frac{h_im_j-m_ih_j-\frac{3}{2}(l_ik_j-k_il_j)}{|z_{ij}|}\notag\\
                                   &=&\sum_{i=1}^n\sum_{j=1(j\not=i)}^n\frac{(h_ik_j-h_jk_i)^3}{2|z_{ij}|}\notag\\
                                   &=&0, \label{eq:c2sum}
\end{eqnarray}
where we have used Eqs.~(\ref{eq:l0}) and (\ref{eq:m0}) in the second equality and the antisymmetry for $i$ and $j$  in the last summation. 
Thus, the bubble equations $h_ic_{2(i)}=0\ (i=1,\ldots,n)$ give $(n-1)$ independent constraint equations for the parameters $(k_i,z_i)\ (i=1,\ldots,n)$.

\subsubsection{Axis}\label{sec:axis}

The $z$-axis of ${\mathbb E}^3$ (i.e., $x=y=0$) in the Gibbons-Hawking space consists of the $(n+1)$ intervals: $I_-=\{(x,y,z)|x=y=0,  z<z_1\}$, $I_i=\{(x,y,z)|x=y=0,z_i<z<z_{i+1}\}\ (i=1,...,n-1)$ and $I_+=\{(x,y,z)|x=y=0,z>z_n\}$. On the $z$-axis, the $1$-forms $\hat\omega_{ij}$ and $\tilde\omega_i$  are, respectively, simplified to 
\begin{eqnarray}
\hat \omega_{ij}=\frac{(z-z_i)(z-z_j)}{z_{ji}|z-z_i||z-z_j|}d\phi,\qquad \tilde\omega_i=\frac{z-z_i}{|z-z_i|}d\phi.
\end{eqnarray}
In particular, on $ I_\pm$, $\hat\omega_{ij}$ and $\tilde\omega_i$ become, respectively,
\begin{eqnarray}
\hat \omega_{ij}=\frac{1}{z_{ji}}d\phi,\qquad \tilde\omega_i=\pm d\phi.
\end{eqnarray}
Hence,  on $I_\pm$, $\hat\omega=\hat\omega_\phi d\phi$ vanishes since
\begin{eqnarray}
\hat\omega&=&\sum_{k,j(k\not=j)}\left(h_km_j+\frac{3}{2}k_kl_j\right)\hat\omega_{kj}-\sum_{j}\left(m_0h_j+\frac{3}{2}k_j\right)\hat\omega_j+c d\phi\notag\\
&=&\sum_{k,j(k\not=j)}\left(h_km_j+\frac{3}{2}k_kl_j\right)\frac{d\phi}{z_{jk}}\mp\sum_{j}\left(m_0h_j+\frac{3}{2}k_j\right)d\phi -\sum_{k,j(k\not=j)}\left(h_km_j+\frac{3}{2}k_kl_j\right)\frac{d\phi}{z_{jk}} \notag\\
&=&\mp\sum_{j}\left(m_0h_j+\frac{3}{2}k_j\right)d\phi \notag\\
&=&\mp\left( m_0+\frac{3}{2}\sum_jk_j \right) d\phi\notag\\
&=&0,
\end{eqnarray}
where we have used Eq.~(\ref{eq:m0}) in the last equality.

On $z\in I_i\ (i=1,...,n-1)$, the 1-forms $\hat\omega_{ij}$ and $\hat\omega_{j}$ are written as
\begin{eqnarray}
\hat\omega_{kj}&=&\frac{z_{ik}z_{ij}}{z_{jk} |z_{ik}z_{ij}|}d\phi\ (k,j\not=i),\qquad \hat\omega_{ij}=-\frac{1}{|z_{ij}|}d\phi\ (j\not=i)\\
\tilde \omega_{j}&=&\frac{z_{ij}}{|z_{ij}|}d\phi\ (j\not=i), \quad 
\qquad \tilde\omega_{i} =d\phi,
\end{eqnarray}
and therefore,
\begin{eqnarray}
\hat\omega_{\phi}-\hat\omega_{0(i)}
&=&-\sum_{j(j\not=i)} \frac{h_im_j-h_jm_i+\frac{3}{2}(k_il_j-k_jl_i) }{z_{ji}}-\left( m_0h_i+\frac{3}{2}k_i\right)
\notag \\
                                      &=&h_ic_{2(i)}\notag \\
                    &=&0,                 
\end{eqnarray}
where we have used Eq.~(\ref{eq:condition3b}) and (\ref{eq:omega}). 
Thus, we can show that  $\hat\omega=0$ also holds on $I_i$ for $i=1,...,n-1$.
We therefore conclude  that  $\hat\omega=0$ holds at each interval $I_\pm$ and $I_i\ (i=1,\ldots,n-1)$. 
This means that there are no Dirac-Misner strings in the spacetime, which can be obtained as the result of the bubble equations~ (\ref{eq:condition3b}) (see \cite{Bena:2007kg,Gibbons:2013tqa}).  

\medskip

Next, we  show the absence of orbifold singularities. 
On the intervals $I_\pm$, the 1-form $\chi$ becomes 
\begin{eqnarray}
\chi&=&\pm d\phi,
\end{eqnarray} 
and on the intervals $I_i\ (i=1,\ldots,n-1)$, it takes the form 
\begin{eqnarray}
\chi
      &=&\left(\sum_{j=1}^i h_j\frac{z-z_j}{|z-z_j|}+\sum_{j=i+1}^n h_j\frac{z-z_j}{|z-z_j|}\right)d\phi \notag \\
      &=&\left(\sum_{j=1}^ih_j-\sum_{j=i+1}^nh_j\right)d\phi.
\end{eqnarray}
The two-dimensional $(\phi,\psi)$-part of the metric on the intervals $I_\pm$ and $I_i$
can be written in the form
\begin{eqnarray}
ds^2_2=(-f^2\omega_\psi^2+f^{-1}H^{-1})(d\psi+\chi_\phi d\phi)^2. \label{eq:axis}
\end{eqnarray}
Here let us use the coordinate basis vectors $(\partial_{\phi_1},\partial_{\phi_2})$ with $2\pi$ periodicity, 
instead of $(\partial_\phi,\partial_\psi)$, where  the coordinates  $\phi_1$ and $\phi_2$ are defined by $\phi_1:=(\psi+\phi)/2$ and $\phi_2:=(\psi-\phi)/2$.  
 It can be shown from~(\ref{eq:axis}) that  the Killing vector $v:=\partial_\phi-\chi_\phi\partial_\psi$ vanishes on each interval. Indeed we can show
\begin{enumerate}
\item on the interval $I_-$, the Killing vector $v_-:=\partial_\phi+\partial_\psi=\partial_{\phi_1}$ vanishes,
\item on each interval $I_i$ ($i=1,...,n-1$), the Killing vector 
\begin{eqnarray}
v_i&:=&\partial_\phi-\chi_\phi \bigr|_{I_i}  \partial_\psi\\
     &=& \frac{1-\chi_{\phi}\bigr|_{I_i}}{2} \partial_{\phi_1}- \frac{1+\chi_{\phi}\bigr|_{I_i}}{2}\partial_{\phi_2}\\
     &=&\frac{1}{2} \left( 1-\sum_{j=1}^ih_j+\sum_{j=i+1}^nh_j \right) \partial_{\phi_1}-\frac{1}{2} \left( 1+\sum_{j=1}^ih_j-\sum_{j=i+1}^nh_j \right) \partial_{\phi_2}\\
     &=&\left(\sum_{j=i+1}^n h_j\right)\partial_{\phi_1}-\left(\sum_{j=1}^i h_j\right)\partial_{\phi_2}
\end{eqnarray}
vanishes, where we have used $\sum_ih_j=1$ in the last equation.
\item on the interval $I_+$, the Killing vector $v_+:=\partial_\phi-\partial_\psi=-\partial_{\phi_2}$ vanishes.

\end{enumerate}

From these, we can observe that the Killing vectors $v_\pm,\ v_i$ on the intervals satisfy 
with

\begin{eqnarray}
{\rm det}\ (v_-^T,v_{1}^T)=h_1,\quad {\rm det}\ (v_{n-1}^T,v_{+}^T)=-h_n, \label{noorbifold1}
\end{eqnarray}
\begin{eqnarray}
{\rm det}\ (v_{i-1}^T,v_{i}^T)&=&-\left(\sum_{j=i}^n h_j\right)\left(\sum_{j=1}^i h_j\right)+\left(\sum_{j=i+1}^n h_j\right)\left(\sum_{j=1}^{i-1} h_{j}\right)\notag\\
                                        &=&-\left(\sum_{j=i}^n h_j\right)\left(\sum_{j=1}^i h_j\right)+\left(\sum_{j=i}^n h_j-h_i\right)\left(\sum_{j=1}^{i} h_{j}-h_i\right)\notag\\
                                        &=&h_i^2- \left(  \sum_{j=i}^n h_j+\sum_{j=1}^i h_j    \right)h_i\notag\\
                                        &=&h_i^2- \left(  \sum_{j=i}^n h_j +h_i   \right)h_i\notag\\
                                        &=&- \left(  \sum_{j=i}^n h_j \right)h_i \notag\\
                                        &=&-h_i. \label{noorbifold2}
\end{eqnarray}
Therefore, it turns out that $|{\rm det}\ (v_-^T,v_{1}^T)|=|{\rm det}\ (v_{n-1}^T,v_{+}^T)|=|{\rm det}\ (v_{i-1}^T,v_{i}^T)|=1$ hold, which means that there exist no orbifold singularities at adjacent intervals, as proved in Ref.~\cite{Hollands:2007aj}.

\medskip

\subsection{Gauge freedom}
As discussed in Ref.~~\cite{Bena:2005ni},  the supersymmetric solutions have a gauge freedom, which means that  for the linear transformation for the harmonic functions $H,K,L$ and $M$, 
\begin{eqnarray}
K\to K+\bar \lambda H,\qquad L\to L-2\bar \lambda K-\bar \lambda^2H,\qquad M\to M-\frac{3}{2}\bar \lambda L+\frac{3}{2}\bar \lambda^2K+\frac{1}{2}\bar \lambda^3H,\label{eq:trans}
\end{eqnarray}
the metric and Maxwell field are invariant, 
where $\bar \lambda$ is a constant. 
Indeed, it is easy to show that under the transformation~(\ref{eq:trans}), $(f, \omega_\psi, \chi$) remain invariant, and the 1-form $\xi$ changes as $\xi\to \xi-\bar \lambda \chi$, which merely corresponds to the gauge shift of $A$, $A\to A+\bar \lambda  d \psi$.  
Using this gauge transformation and the appropriate choice of $\bar \lambda$,  one can set 
\begin{eqnarray}
k_m=0, \label{eq:km0}
\end{eqnarray}
for a certain $m\ (m=1,\ldots,n)$ because the coefficient of $1/r_m$ in $K$ changes $k_m\to  k_m+\bar\lambda h_m$.
Moreover, using the shift of $z\to z+$const., one can set
\begin{eqnarray}
z_m=0 \label{eq:zm0}
\end{eqnarray}
for a certain $m\ (m=1,\ldots,n)$.

\subsection{Parameter counting}
The solution~(\ref{eq:solution_metric}) and (\ref{eq:solution_1form})  includes the $4n+3$ continuous parameters $(k_i,l_0,l_i,m_0,m_i,z_i,c)$ and the $n$ discrete parameters $h_i=\pm 1\ (i=1,\ldots n)$.
The conditions~(\ref{eq:l0}), (\ref{eq:m0}), (\ref{eq:c})
(\ref{eq:condition1}), (\ref{eq:condition2}), (\ref{eq:condition3b}) and the gauge conditions~(\ref{eq:km0}), (\ref{eq:zm0}) reduce  the number of independent  continuous parameters from $4n+3$ to $n-1$, where the bubble equations~(\ref{eq:condition3b}) give not $n$ but rather $(n-1)$ independent equations due to the constraint equation~(\ref{eq:c2sum}), and the condition~(\ref{eq:sumhi}) reduces the number of independent  discrete parameters from $n$ to $n-1$. Moreover, these parameters must be subject to the $n$ inequalities (\ref{eq:c1ineq}).  

\medskip
It follows from the constraint equation~(\ref{eq:sumhi}) and the conditions~(\ref{eq:hi}) that the number $n$ of centers ${\bm r}={\bm r}_i$ must be odd, and  so in Sec.~\ref{sec:reflection}, we consider three centers and five centers as the simplest nontrivial examples of the microstate geometries (the case $n=1$ corresponds to Minkowski spacetime).

\section{Physical properties}
\label{sec:analysis}
Under the appropriate boundary conditions mentioned in the previous section,  let us investigate some physical properties of the solutions.

\subsection{Conserved quantities}

To begin with, we consider conserved quantities of the microstate geometries. 
From the boundary conditions at infinity~(\ref{eq:l0})-(\ref{eq:c}), the ADM mass and two ADM angular momenta can be  computed as
\begin{eqnarray}
M&=&\frac{\sqrt{3}}{2}Q=3\pi\left[\left(\sum_{i}k_i\right)^2\left(\sum_ih_i \right)^{-1}+\sum_{i}l_i\right], \label{eq:mass}\\
J_\psi&=&\pi \left[ \left(\sum_{i}k_i\right)^3+\sum_{i}m_i+\frac{3}{2}\left(\sum_ih_i \right)^{-1}\left(\sum_{i}k_i\right)\left(\sum_{i}l_i\right)\right], \label{eq:jpsi}\\
J_\phi&=&\frac{3\pi}{2}\left(\sum_i h_i\right)^{-1}\left[-\left(\sum_{i}k_i\right) \left(\sum_{i}h_iz_i\right)+\left(\sum_{i }k_iz_i\right)  \right], \label{eq:jphi}
\end{eqnarray}
where $Q$ is the electric charge, which saturates the BPS bound~\cite{Gibbons:1993xt}.

Each interval $I_i$ ($i=1,...,n-1$),  which is introduced in Sec.~\ref{sec:axis}, denotes the bubble which is topologically a two-dimensional sphere since  
the $\psi$-fiber of the Gibbons-Hawking space~(\ref{eq:GH}) collapses to zero at the centers $z=z_i$ and $z=z_{i+1}$, and so along the interval, the fiber sweeps out two-dimensional sphere. 
Since the Maxwell gauge field $A_\mu$ is obviously smooth on the bubbles, the magnetic fluxes through $I_i$ ($i=1,...,n-1$) can be defined as 
\begin{eqnarray}
q[I_i]:=\frac{1}{4\pi}\int_{I_i}F,
\end{eqnarray}
which are computed as
\begin{eqnarray}
q[I_i]=[-A_\psi]^{z=z_{i+1}}_{z=z_i}=\frac{\sqrt{3}}{2}\left( \frac{k_i}{h_i}-\frac{k_{i+1}}{h_{i+1}}\right)~~ (i=1,..., n-1).
\label{eq:fluxes}
\end{eqnarray}

\subsection{Evanescent ergosurface}
\begin{figure}[H]
\centering
\includegraphics[width=6cm]{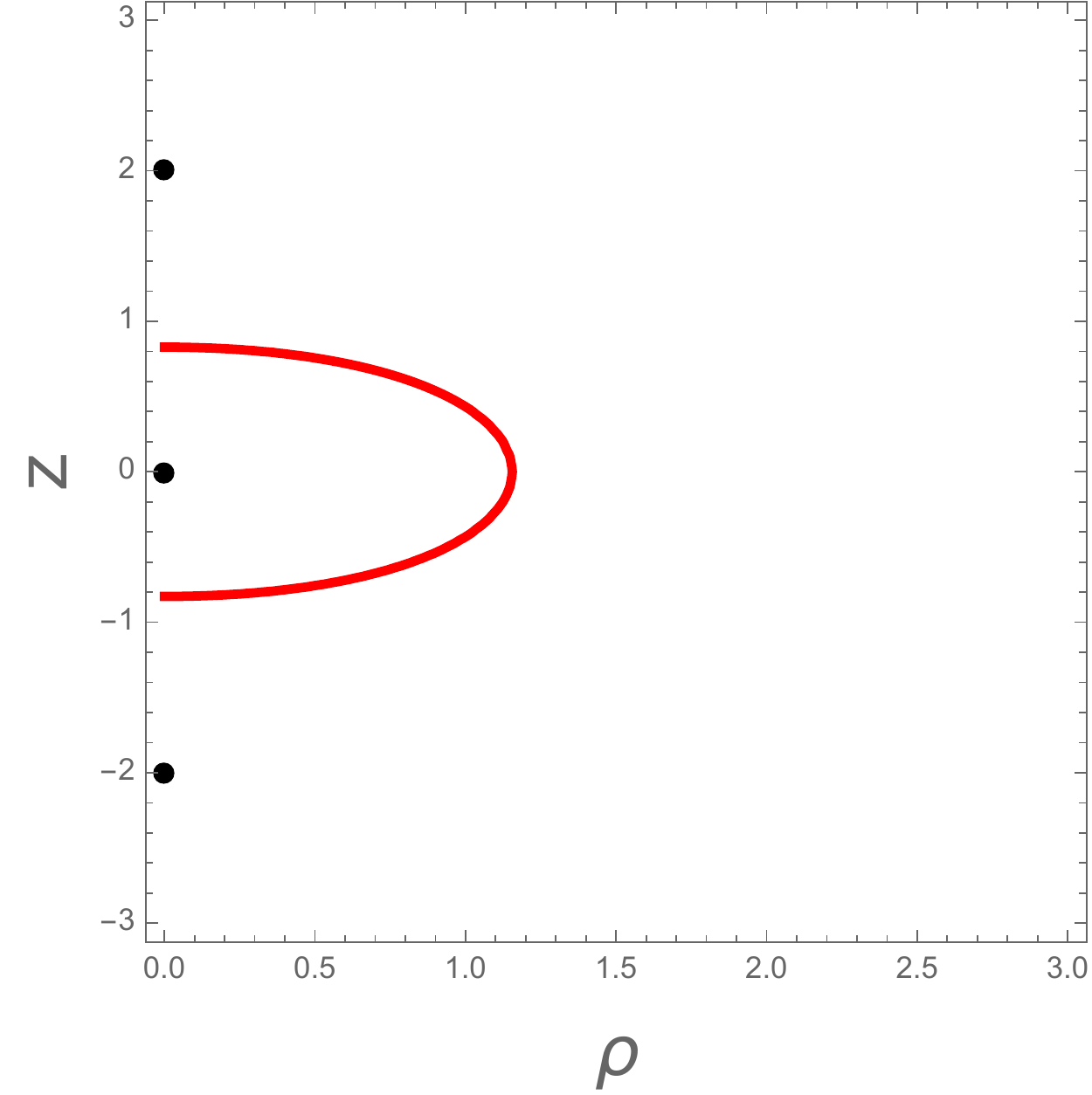}

\caption{Evanescent ergosurface in the microstate geometry for $n=3$ in the $(\rho,z)$-plane: 
The black points corresponds to three centers that are located at ${\bm r_1}$, ${\bm r_2}$ and ${\bm r_3}$ on the $z$-axis, and 
the red curve denotes an evanescent ergosurface, which surrounds a center at ${\bm r_2}=(0,0)$ but does not other two centers ${\bm r_1}=(0,-2)$ and ${\bm r_3}=(0,2)$.}

\label{fig:sp_1}
\end{figure}

The existence of ergoregions gives rise to strong instability due to a superradiant-triggered mechanism in spite of the existence of the horizon~\cite{Cardoso:2014sna,Brito:2015oca}. 
It was demonstrated that a certain class of non-supersymmetric microstate geometries with ergoregion in type IIB supergravity are unstable, which is a general feature of horizonless geometries with ergoregion~\cite{Cardoso:2005gj}. 
The BPS microstate geometries does not admit the presence of ergoregions but evanescent ergosurfaces~\cite{Gibbons:2013tqa,Niehoff:2016gbi}, which are defined as timelike hypersurfaces such that a stationary Killing vector field becomes null there and timelike everywhere except there. 
Reference~\cite{Eperon:2016cdd} proved that on such surfaces, massless particles with zero energy ($E=0$) relative to infinity move along stable trapped null geodesics. 
Since this stably trapping leads to a classical non-linear instability of the spacetime~\cite{Keir:2014oka,Cardoso:2014sna,Eperon:2016cdd}, it is of physical importance to investigate the existence of evanescent ergosurfaces, which exist at $f=0$ which corresponds to 
\begin{eqnarray}
H=\sum_{i=1}^n\frac{h_i}{r_i}=0.
\end{eqnarray}
For simplicity, let us consider the microstate geometries with reflection symmetry 
$z_{m}=-z_{n-m+1}$ and $k_{m}=k_{n-m+1}$\ ($m=1,\ldots n$).
For the microstate geometries with three centers ($n=3$) and $(h_1,h_2,h_3)=(1,-1,1)$, they intersect the $z$-axis at the points
\begin{eqnarray}
F_3(z):=|z||z-z_3|-|z-z_1||z-z_3|+|z||z-z_1|=0.
\end{eqnarray}
It turns out from simple computations that $F_3(z)=0$ has no root on $I_\pm$ and a single root $I_i\ (i=1,2)$. 
As seen FIG.~\ref{fig:sp_1}, the evanescent ergosurfaces on the timeslice $t=$const. is the closed surface surrounding the center ${\bm r}_2=(0,0,0)$,  where we have introduced the radial coordinate by $\rho=\sqrt{x^2+y^2}$.

\begin{figure}[H]
\begin{tabular}{lll}
\begin{minipage}[t]{0.3\hsize}
\centering
\includegraphics[width=5cm]{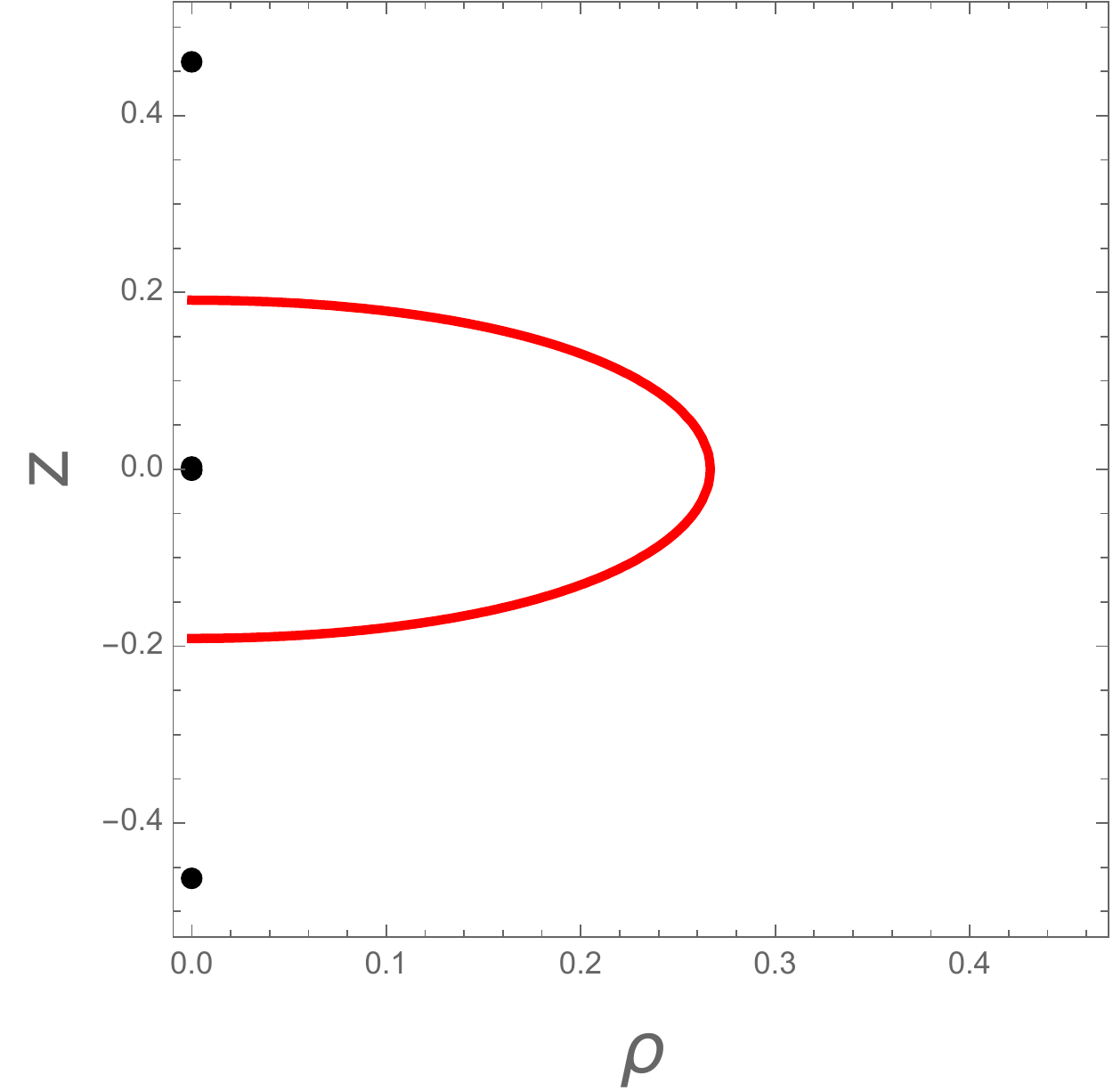}
 \end{minipage} &
 
  \begin{minipage}[t]{0.3\hsize}
 \centering
\includegraphics[width=5cm]{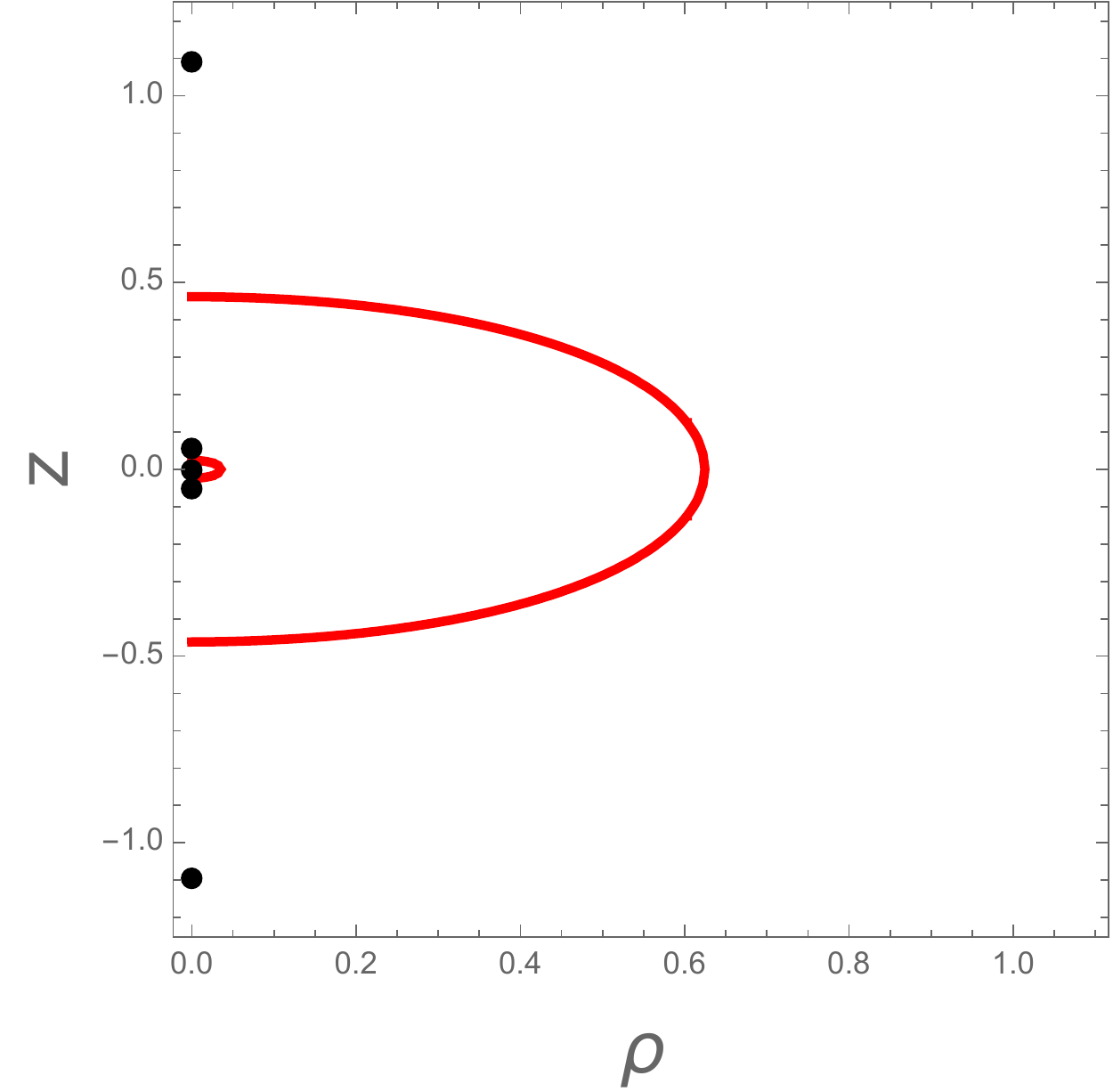}
  \end{minipage} &
 
  \begin{minipage}[t]{0.3\hsize}
 \centering
\includegraphics[width=5cm]{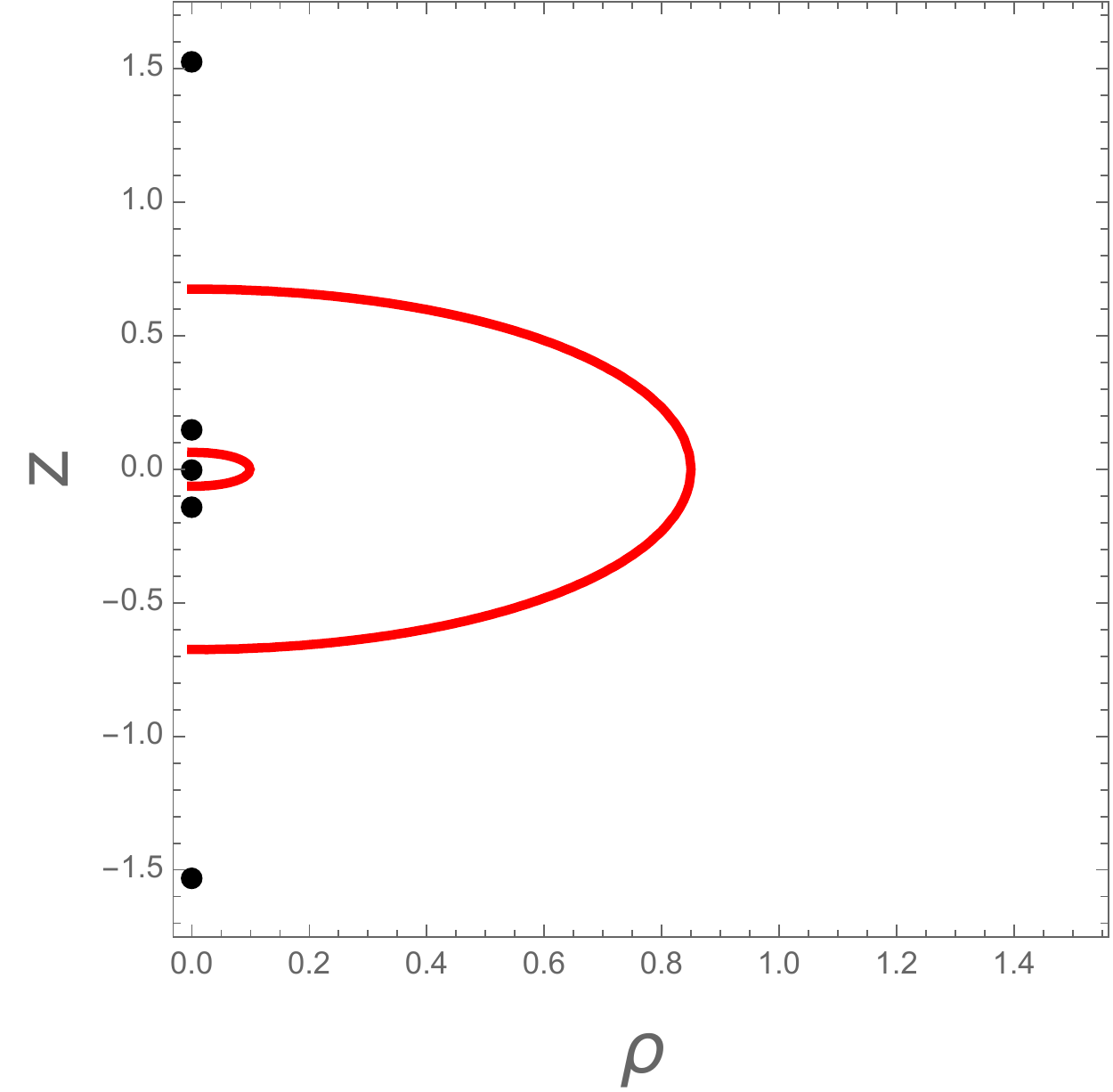}
 \end{minipage}\\

 \begin{minipage}[t]{0.3\hsize}
\centering
\includegraphics[width=5cm]{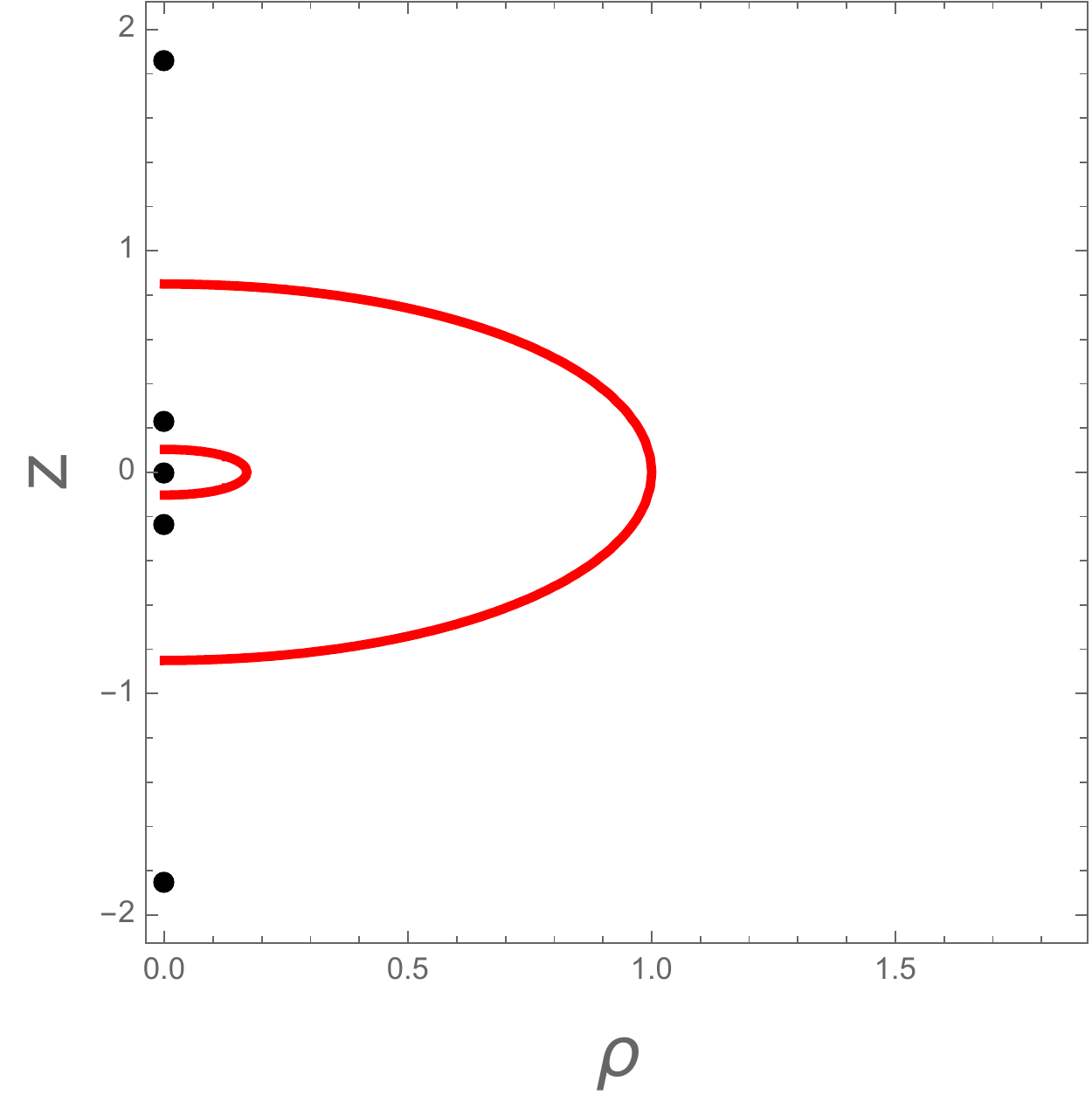}
 \end{minipage} &
 
  \begin{minipage}[t]{0.3\hsize}
 \centering
\includegraphics[width=5cm]{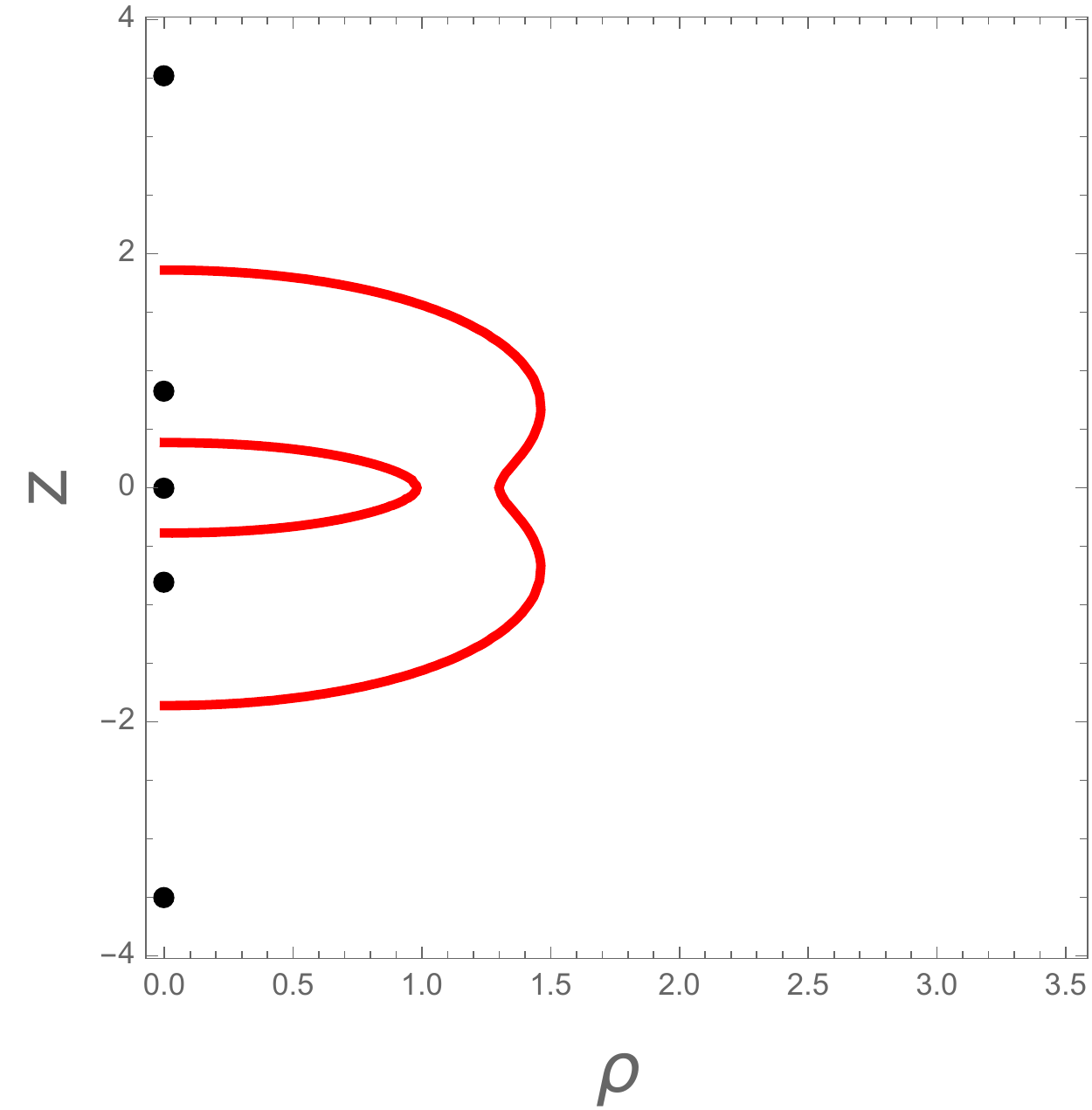}
  \end{minipage} &
 
  \begin{minipage}[t]{0.3\hsize}
 \centering
\includegraphics[width=5cm]{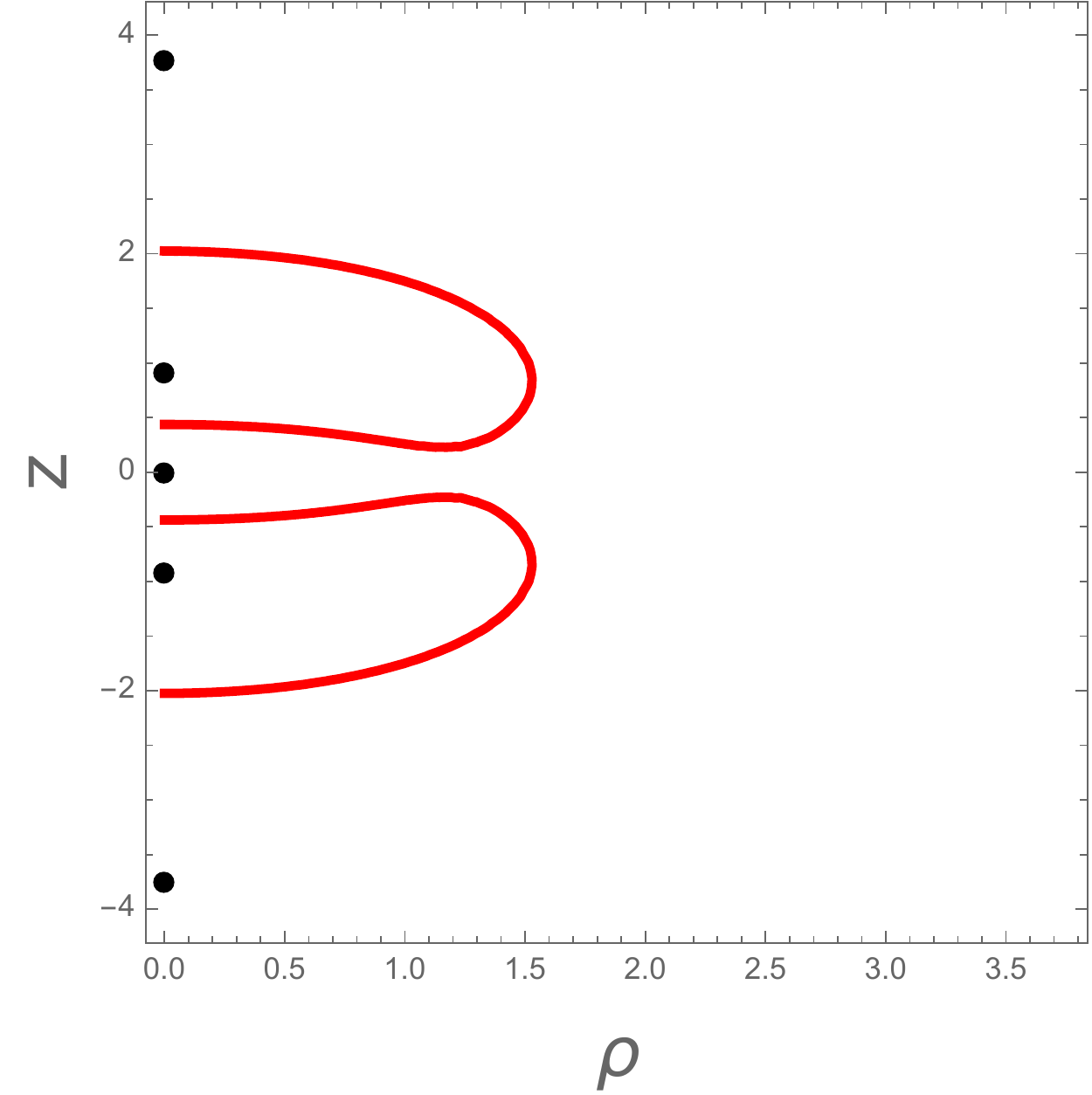}
 \end{minipage}

 \end{tabular} 
 \caption{Evanescent ergosurfaces in the microstate geometry with five centers in the $(\rho,z)$-plane for $k_3=0, \ k_4=k_2,\ k_5=k_1,\ z_3=0,\ z_4=-z_2,\ z_5=-z_1$: 
 The upper and lower figures correspond to the ratios $k_2/k_1=0.1,\ 0.6,\ 0.9,$ and $k_2/k_1=1.1,\ 1.9,\ 2.0$, respectively, from left to right.
The black points correspond to the five centers that are located at ${\bm r_i}\ (i=1,\ldots,5)$ on the $z$-axis, and 
the red curves denote the evanescent ergosurfaces, whose shapes depend on $k_1$ and $k_2$. 
}

 \label{fig:EESd5}
\end{figure}

\medskip
For the microstate geometry with five centers ($n=5$) and $(h_1,h_2,h_3,h_4,h_5)=(1,-1,1,-1,1)$, they intersect the $z$-axis at the points $z$ satisfying $F_5(z)=0$, where $F_5(z)$ is written as 
\begin{eqnarray}
F_5(z):&=&|z+z_2||z||z-z_2||z-z_1|-|z+z_1||z||z-z_2||z-z_1| +|z+z_1||z+z_2||z-z_2||z-z_1| \notag\\
      &&-|z+z_1||z+z_2||z||z-z_1|+ |z+z_1||z+z_2||z||z-z_2|.
\end{eqnarray}
The roots of the equation $F_5(z)=0$ are determined by the ratio $k_2/k_1$ through the bubble equations~(\ref{eq:condition3b}). 
As seen in FIG.~\ref{fig:EESd5},  for the small ratio $0<k_2/k_1\ll1$, the intervals $z_{21}(z_{54})$  of $I_1\ (I_4)$ are much larger the intervals $z_{32}(z_{43})$ of $I_2\ (I_3)$, whereas for the comparable ratio $1 \lesssim k_2/k_1 \lesssim 2$,  the intervals $z_{21}(z_{54})$ also become comparable with $z_{32}(z_{43})$. 
This reason can be physically interpreted as the result that the magnetic fluxes need  to support the bubbles. 
More precisely,  this is caused by the force balance between a gravitational force that tend to contract the bubbles and a repulsive force by the magnetic fluxes that tend to expand the bubbles. 
For $k_2/k_1\ll1$, the magnetic flux through $I_1\ (I_4)$ is much larger than one through $I_2\ (I_3)$ [$|q[I_1]| \gg |q[I_2]|\ (|q[I_4]| \gg |q[I_3]|)$], and so the size of the bubble on $I_1\ (I_4)$ is larger than $I_2\ (I_3)$, whereas for $k_2/k_1\simeq 2$, two magnetic fluxes are comparable [$|q[I_1]|\simeq |q[I_2]|\ (|q[I_4]|\simeq |q[I_3]|)$], and hence the size of the bubbles also becomes comparable. 
For $k_2/k_1\ll1$, the evanescent ergosurface exists as a common surface surrounding three centers ${\bm r}={\bm r}_i=(0,0,z_i)\ (i=2,3,4)$, for $k_2/k_1\simeq 1$, another ergosurface appears as the surface surrounding the center ${\bm r}={\bm r}_3=(0,0,z_3)$, whereas for $k_2/k_1\simeq 2$, two ergosurfaces combine into one,  and thereafter separates into two parts.

\section{Microstate geometries with reflection symmetry}\label{sec:reflection}

 In Sec.~\ref{sec:msg}, we have considered the stationary and bi-axisymmetric microstate geometries with $n$ centers  on the $z$-axis of the Gibbons-Hawking space which satisfy the bubble equations~(\ref{eq:condition3b}).
The nasty constraint equations (for the parameters included in the solutions) make it difficult for us to understand the physical properties.
In this section, in addition to such symmetry assumptions, we impose a further reflection symmetry on the solutions: 
\begin{eqnarray}
z_{m}=-z_{n-m+1},\qquad k_{m}=k_{n-m+1}\ (m=1,\ldots n), 
\end{eqnarray}
which means the invariance of the solutions under the transformation $z\to - z$.
This additional assumption extremely simplifies the bubble equations so that one can solve them and express $z_i\ (1,\ldots,n)$ in terms of $k_i\ (i=1,\ldots,n)$, at least, for small $n$. 
In particular, it is easy to show from Eq.~(\ref{eq:jphi}) that the angular momentum $J_\phi$ always vanishes under the additional symmetry assumption. 
In this section, for simplicity, let us consider only two cases of $n=3$ and $n=5$.

\subsection{Three-center solution}
First, let us consider the solution with three centers ($n=3$) and $(h_1,h_2,h_3)=(1,-1,1)$ that describes the simplest asymptotically flat, stationary and bi-axisymmetric microstate geometry, which has the four parameters $(k_1,k_3,z_1,z_3)$, where we have set $k_2=0$ and $z_2=0$ from the two gauge conditions~(\ref{eq:km0}) and (\ref{eq:zm0}). 
Moreover, under the assumption of the reflection symmetry 
\begin{eqnarray}
z_3=-z_1=:a\ (>0),\qquad k_3=k_1, \label{eq:reflection_n3}
\end{eqnarray}
the bubble equations~(\ref{eq:condition3b}) are simply written as 
\begin{eqnarray}
c_{2(1)}=-\frac{1}{2}c_{2(2)}=c_{2(3)}=\frac{k_1[k_1^2-3a]}{2a}=0, \label{eq:c2_n3}
\label{eq:c23}
\end{eqnarray}
which imply
\begin{eqnarray}
&&k_1=0,\\
&&a=\frac{k_1^2}{3}.   \label{eq:c2l}
\end{eqnarray}
It is obvious that in the former case $h_ic_{1(i)}=0\ (i=1,2,3)$, and so the inequalities~(\ref{eq:c1ineq}) cannot be satisfied. 
In the meanwhile, in the latter case, the inequalities~(\ref{eq:c1ineq}) can be automatically satisfied because $h_ic_{1(i)}\ (i=1,2,3)$ can be directly computed as
\begin{eqnarray}
h_1c_{1(1)}=h_3c_{1(3)}=4,\quad h_2c_{1(2)}=5.
\end{eqnarray}
Therefore, for arbitrary nonzero $k_1$, this describes a regular and causal solution of an asymptotically flat, stationary microstate geometry with the bi-axisymmetry and reflection symmetry. 
This solution was previously analyzed in Ref.~\cite{Gibbons:2013tqa}. 

\begin{figure}[H]
\centering
\includegraphics[width=8cm]{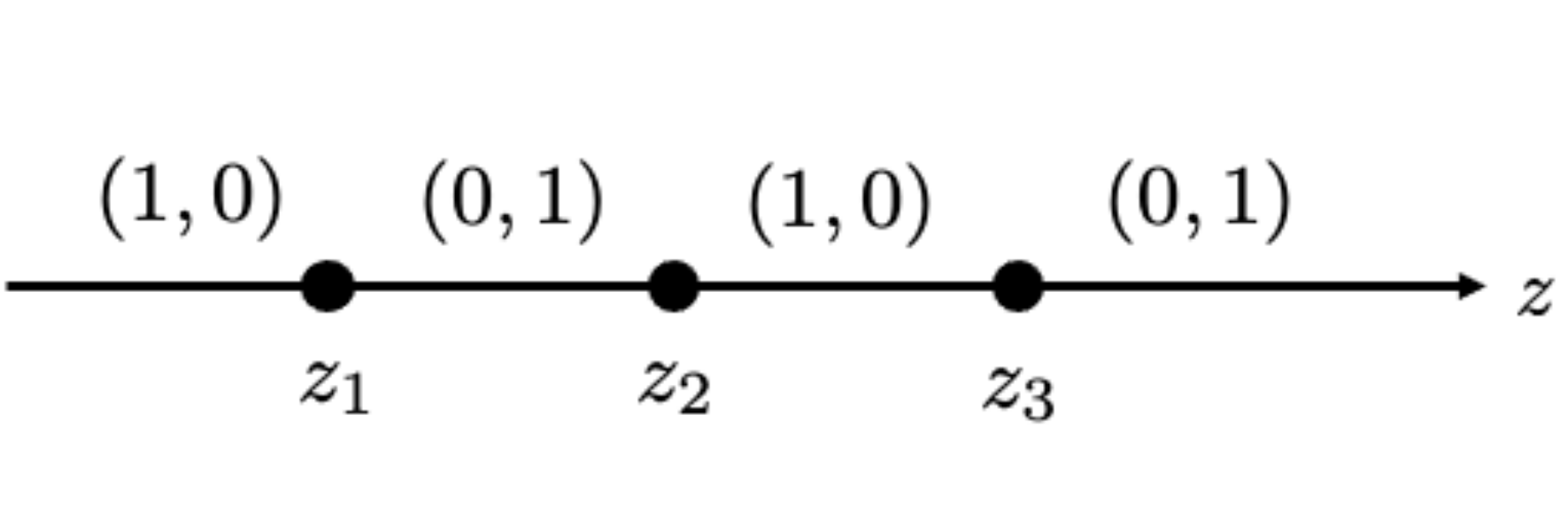}
\caption{Rod structure for the microstate geometry with three centers and $(h_1,h_2,h_3)=(1,-1,1)$.}
\label{fig:rod_n3}
\end{figure}

\medskip
The $z$-axis of ${\mathbb E}^3$ in the Gibbons-Hawking space consists of the four intervals: 
$I_-=\{(x,y,z)|x=y=0,  z<z_1\}$, $I_i=\{(x,y,z)|x=y=0,z_i<z<z_{i+1}\}\ (i=1,2)$ and $I_+=\{(x,y,z)|x=y=0,z>z_3\}$. 
From the result in Sec.~\ref{sec:axis}, one can see 
\begin{enumerate}
\item on $I_-$, the Killing vector $v_-=\partial_{\phi_1}$ vanishes,
\item on $I_1$, the Killing vector 
$
v_1=\left(\sum_{j=2}^3 h_j\right)\partial_{\phi_1}-h_1\partial_{\phi_2}=-\partial_{\phi_2}
$
vanishes,

\item on $I_2$, the Killing vector 
$
v_2=h_3 \partial_{\phi_1}-\left(\sum_{j=1}^2 h_j\right)\partial_{\phi_2}=\partial_{\phi_1}
$
vanishes, and

\item on $I_+$, the Killing vector $v_+=-\partial_{\phi_2}$ vanishes.

\end{enumerate}
 Thus the rod structure of this three-center microstate geometry is displayed in Fig.~\ref{fig:rod_n3}.

\medskip
 Under the symmetric conditions~(\ref{eq:reflection_n3}) and gauge conditions $k_2=0$, $z_2=0$, the ADM mass and two ADM angular momenta in Eqs.~(\ref{eq:mass})-(\ref{eq:jphi}) are reduced to
\begin{eqnarray}
M&=&\frac{\sqrt{3}}{2}Q=6\pi k_1^2,\label{eq:mass_n3}\\
J_\psi&=&3\pi k_1^3,\label{eq:jpsi_n3}\\
J_\phi&=&0,\label{eq:jphi_n3}
\end{eqnarray}
and the magnetic fluxes in Eq.~(\ref{eq:fluxes}) are written as
\begin{eqnarray}
q[I_1]=-q[I_2]=\frac{\sqrt{3}}{2}k_1.
\end{eqnarray}

\subsection{Five-center solution}
Next, let us consider the stationary, bi-axisymmetric  microstate geometry with five centers ($n=5$), which has the four parameters $(k_1,k_2,z_1,z_2)$ under the reflection-symmetric conditions
\begin{eqnarray}
k_5=k_1,\quad k_4=k_2,\quad z_5=-z_1=:a+b,\quad z_4=-z_2=:b
\end{eqnarray}
and the gauge conditions $k_3=0$, $z_3=0$. 
Here, let us notice that for the five-center solutions,  there are two possible types of reflection-symmetric solutions, one with  $(h_1,h_2,h_3,h_4,h_5)=(1,-1,1,-1,1)$ and one with $(h_1,h_2,h_3,h_4,h_5)=(-1,1,1,1,-1)$, but 
the latter numerically seems not to satisfy the conditions  (\ref{eq:c1ineq}). 
Thus, we here concentrate on only the former, in which case the conditions~(\ref{eq:condition3b}) are simplified  to give
\begin{eqnarray}
&&2h_1c_{2(1)}=2h_5c_{2(5)}=-3(k_1+2k_2)-\frac{k_1^3}{a+b}+\frac{(k_1+k_2)^3}{a}+\frac{(k_1+k_2)^3}{a+2b}=0, \label{eq:c21_n5}\\
&&2h_2c_{2(2)}=2h_4c_{2(4)}=3(2k_1+3k_2)-\frac{k_2^3}{b}-\frac{(k_1+k_2)^3}{a}-\frac{(k_1+k_2)^3}{a+2b}=0, \label{eq:c22_n5}\\
&&h_3c_{2(3)}=-3(k_1+k_2)+\frac{k_1^3}{a+b}+\frac{k_2^3}{b}=0,\label{eq:c23_n5}
\end{eqnarray}
where we note that Eqs.~(\ref{eq:c21_n5})-(\ref{eq:c23_n5}) are not independent due to the constraint equation $\sum_{i=1}^5 h_ic_{2(1)}=2h_1c_{2(1)} +2h_2c_{2(2)}+h_3c_{2(3)}=0$. 
Therefore, this solution has only two independent parameters. 
If we regard $a$ and $b$ as the functions of $k_1$ and $k_2$ from Eqs.~(\ref{eq:c21_n5}), (\ref{eq:c23_n5}),  this solution is a two-parameter family for $(k_1,k_2)$. 

Furthermore, the parameters $k_1$ and $k_2$ must satisfy the inequalities (\ref{eq:c1ineq}), which are reduced to
\begin{eqnarray}
&&h_1c_{1(1)}=h_5c_{1(5)}=1-\frac{k_1^2}{a+b}+\frac{(k_1+k_2)^2}{a}+\frac{(k_1+k_2)^2}{a+2b}>0,\label{eq:c15}\\
&&h_2c_{1(2)}=h_4c_{1(4)}=-1+\frac{k_2^2}{b}+\frac{(k_1+k_2)^2}{a}+\frac{(k_1+k_2)^2}{a+2b}>0,\label{eq:c14}\\
&&h_3c_{1(3)}=1-\frac{2k_1^2}{a+b}+\frac{2k_2^2}{b}>0,\label{eq:c13_n5}
\end{eqnarray}
together with the inequalities
\begin{eqnarray}
a>0,\qquad b>0. \label{eq:ab_positive}
\end{eqnarray}
In the below, we assume $k_1\not=0$ and $k_2\not=0$   
because  from Eqs.~(\ref{eq:c21_n5}) and (\ref{eq:c23_n5}), the case $k_1=0$ leads to
\begin{eqnarray}
(a,b)=\left(\frac{-1\pm\sqrt{5}}{6}k_2^2, \frac{1}{3}k_2^2 \right),
\end{eqnarray}
where  only the solution with the positive sign can satisfy (\ref{eq:c15})-(\ref{eq:ab_positive}) and  has  $j^2=25/24$, and from Eqs.~(\ref{eq:c21_n5}) -(\ref{eq:c23_n5}), the case $k_2=0$ yields $(a,b)=(k_1^2/3,0)$, which cannot satisfy one of the inequalities~(\ref{eq:ab_positive}). 
In what follows, we remove both cases of $k_1=0$ and $k_2=0$.

\begin{figure}[H]
\centering
\includegraphics[width=10cm]{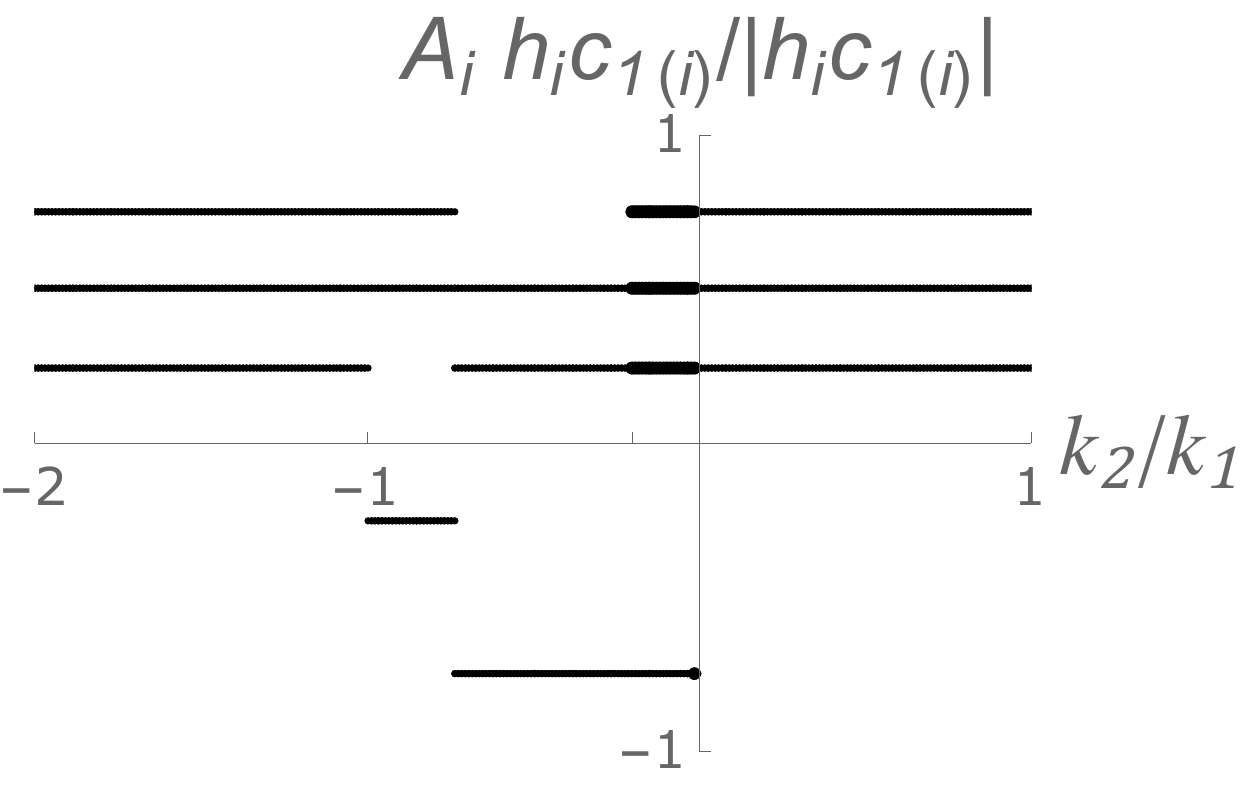}

\caption{The plots of $A_i h_ic_{1(i)}/|h_ic_{1(i)}|$ $[i=1,2,3,\ (A_1,A_2,A_3)=(0.25,0.5,0.75)]$ for the microstate geometry with five centers and $(h_1,h_2,h_3,h_4,h_5)=(1,-1,1,-1,1)$, where we set $k_1=1$. 
The inequalities~(\ref{eq:c15})-(\ref{eq:ab_positive}) are simultaneously satisfied in the range $k_2/k_1<-1, \ -0.2063...<k_2/k_1<0, \ k_2/k_1>0$, where all graphs are positive. 
In particular, in the range $-0.2063...<k_2/k_1<0$,  the solution to Eqs.~(\ref{eq:c21_n5})-(\ref{eq:c23_n5})  has the two branches which have the same nonzero pair of $(k_1,k_2)$ but two different positive pairs of $(a,b)$.  
One of two branches cannot satisfy the inequality~(\ref{eq:c15}). }

\label{fig:parameter_n5}
\end{figure}

 As shown in~Fig.\ref{fig:parameter_n5}, these inequalities are equivalent with 
\begin{eqnarray}
k_2/k_1<-1, \quad  -0.2063...<k_2/k_1<0, \quad  k_2/k_1>0. \label{eq:range}
\end{eqnarray}

\begin{figure}[H]
\centering
\includegraphics[width=8cm]{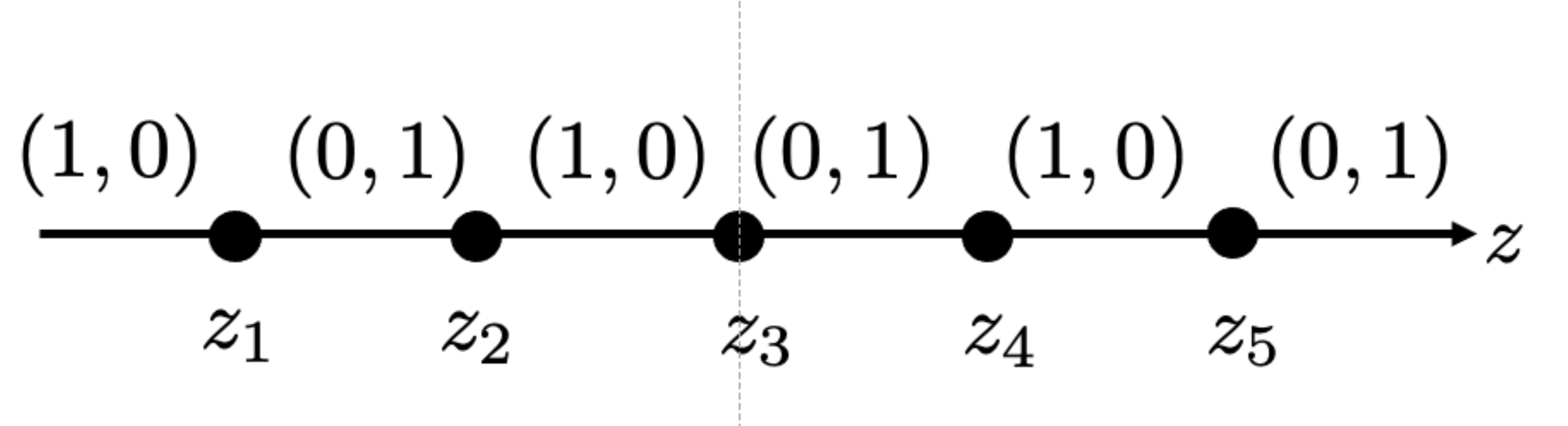}

\caption{Rod structure for the microstate geometry with five centers and $(h_1,h_2,h_3,h_4,h_5)=(1,-1,1,-1,1)$.}

\label{fig:rod_n5}
\end{figure}

The $z$-axis of ${\mathbb E}^3$ in the Gibbons-Hawking space consists of the six intervals: 
$I_-=\{(x,y,z)|x=y=0,  z<z_1\}$, $I_i=\{(x,y,z)|x=y=0,z_i<z<z_{i+1}\}\ (i=1,...,4)$ and $I_+=\{(x,y,z)|x=y=0,z>z_5\}$. 
Applying the result in Sec.~\ref{sec:axis} to this solution, one can see 
\begin{enumerate}
\item on $I_-$, the Killing vector $v_-=\partial_{\phi_1}$ vanishes,

\item on $I_1$, the Killing vector 
$
v_1=\left(\sum_{j=2}^5 h_j\right)\partial_{\phi_1}-h_1\partial_{\phi_2}=-\partial_{\phi_2}
$
vanishes, 

\item on $I_2$, the Killing vector 
$
v_2=\left(\sum_{j=3}^5 h_j\right)\partial_{\phi_1}-\left(\sum_{j=1}^2 h_j\right)\partial_{\phi_2}=\partial_{\phi_1}
$
vanishes,

\item on $I_3$, the Killing vector 
$v_3=\left(\sum_{j=4}^5 h_j\right)\partial_{\phi_1}-\left(\sum_{j=1}^3 h_j\right)\partial_{\phi_2}=-\partial_{\phi_2}$
vanishes,

\item on $I_4$, the Killing vector
$v_4=h_5\partial_{\phi_1}-\left(\sum_{j=1}^4 h_j\right)\partial_{\phi_2}=\partial_{\phi_1}$
vanishes, and 

\item on $I_+$, the Killing vector $v_+=-\partial_{\phi_2}$ vanishes,
\end{enumerate}
Thus, it turns out that this five-center microstate geometry has the rod structure displayed in Fig.~\ref{fig:rod_n5}.

\medskip
For this solution, the ADM mass and two ADM angular momenta in Eqs.~(\ref{eq:mass})-(\ref{eq:jphi}) are reduced to  
\begin{eqnarray}
M&=&\frac{\sqrt{3}}{2}Q=6\pi( k_1^2+4k_1k_2+3k_2^2),\label{eq:mass_n5}\\
J_\psi&=&3\pi (k_1^3+6k_1^2k_2+10k_1k_2^2+5k_2^3),\label{eq:jpsi_n5}\\
J_\phi&=&0, \label{eq:jphi_n5}
\end{eqnarray}
and the magnetic fluxes in Eq.~(\ref{eq:fluxes}) are written as
\begin{eqnarray}
q[I_1]=-q[I_4]=\frac{\sqrt{3}}{2}(k_1+k_2),\quad q[I_2]=-q[I_3]=-\frac{\sqrt{3}}{2}k_2.
\end{eqnarray}

\subsection{Comparison with BMPV black hole} \label{sec:BMPV}
Finally, we compare the BPS microstate geometries for $n=3$ and $n=5$ described in the previous section with the rotating BPS black hole in the five-dimensional minimal supergravity, i.e., the BMPV black hole~\cite{Breckenridge:1996is}, which carries mass (saturated the BPS bound) and equal angular momenta ($J_\phi=0$).
For this purpose, let us define a dimensionless angular momentum by
\begin{eqnarray}
j:=\frac{3\sqrt{3\pi} |J_\psi|}{M^{3/2}}.
\end{eqnarray}
For the BMPV black hole, 
the dimensionless angular momentum $j$ has the range of
\begin{eqnarray}
0\le j< 1,
\end{eqnarray}
where $j=0$ corresponds to the extremal Reissner-Nordstrom black hole. 
The absence of CTCs around the horizon requires the upper bound,  $j=1$. 

It is shown from Eqs.~(\ref{eq:mass_n3}) and (\ref{eq:jpsi_n3}) that for $n=3$, the squared angular momentum $j^2$ takes only the value of
\begin{eqnarray}
j^2=\frac{9}{8}\ (>1),
\end{eqnarray}
which is a larger value than the upper bound for the BMPV black hole.

\begin{figure}[h]

\begin{tabular}{cc}

\begin{minipage}[t]{0.5\hsize}
\centering
\includegraphics[width=8cm]{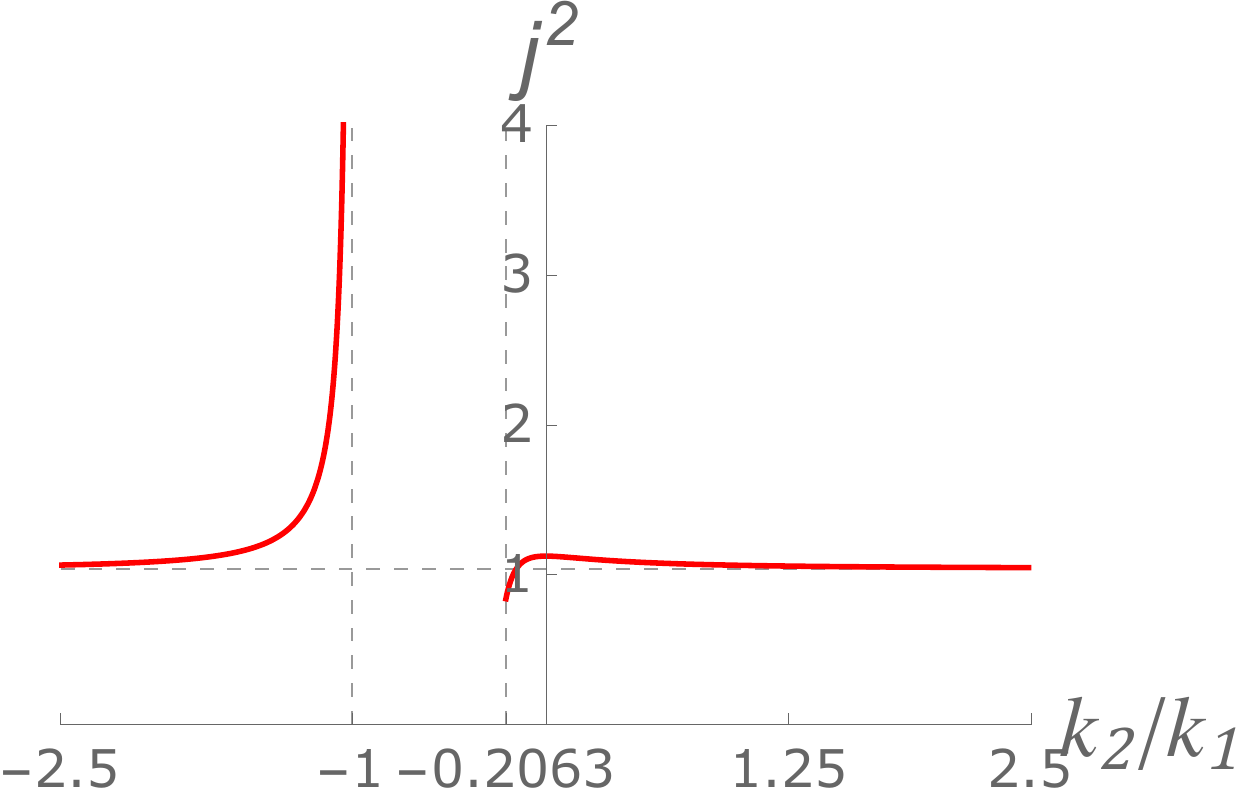}
 \end{minipage} &
 
  \begin{minipage}[t]{0.5\hsize}
 \centering
\includegraphics[width=8cm]{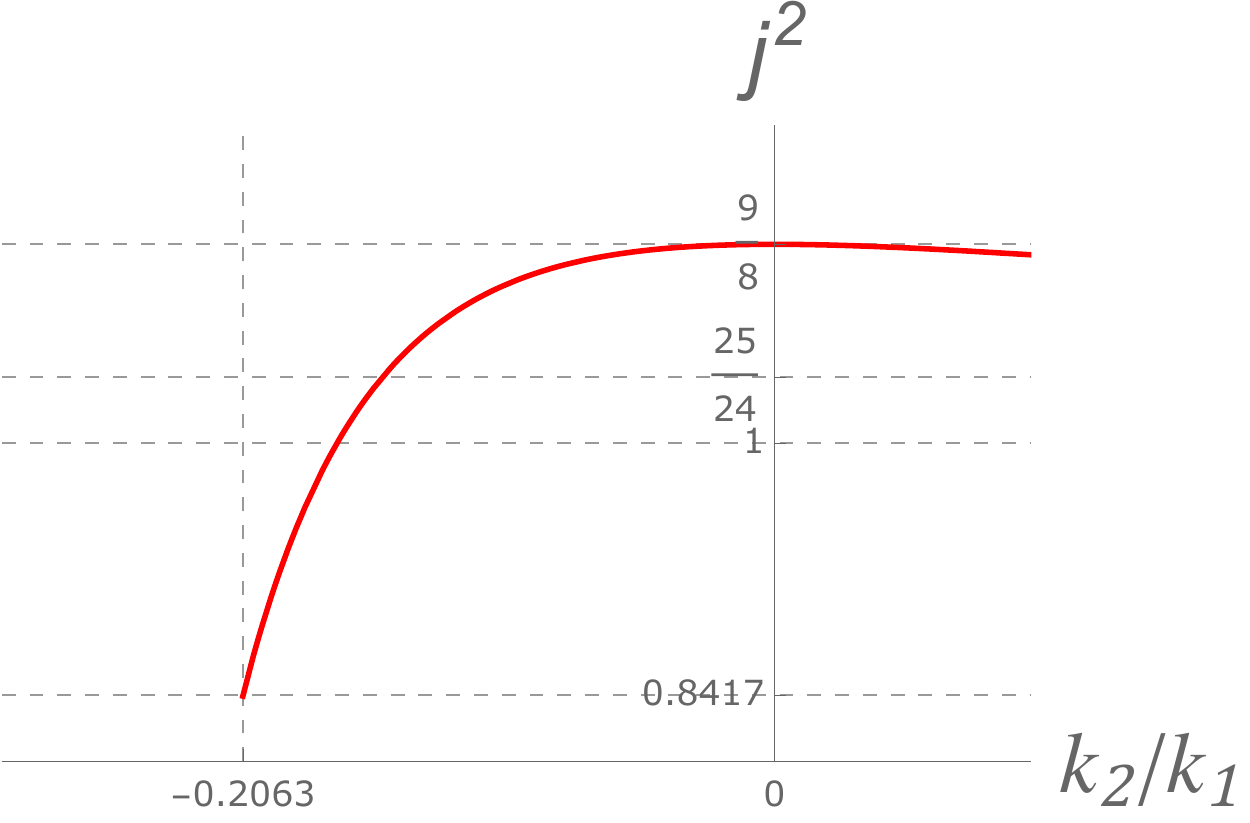}
  \end{minipage}
  
\end{tabular} 
    
\caption{The range of $j^2$ for the asymptotically flat, stationary, bi-axisymemtric and  reflection-symmemtic microstate geometry with five centers ($n=5$). The left figure shows the plots of $j^2$, and the right figure  the close-up region of $-0.206....< k_2/k_1<0$ in the left figure.}
 \label{fig:MvsJ}
\end{figure}

\medskip
Similarly, for $n=5$, we evaluate the value of the squared angular momentum $j^2$  from  Eqs.~(\ref{eq:mass_n5}) and (\ref{eq:jpsi_n5}), where the ratio $k_2/k_1$ lies in the range~(\ref{eq:range}).
As seen in Fig.~\ref{fig:MvsJ},  The  squared angular momentum $j^2$ asymptotically approaches $25/24$ at $k_2/k_1\to -\infty$.
For $k_2/k_1<-1$,  $j^2$ monotonically increases and diverges at $k_2/k_1\to -1$, whereas 
for $k_2/k_1>-1$, it has the lower bound $0.841...$ at $k_2/k_1\to -0.206...$, where Eqs.~(\ref{eq:c21_n5})-(\ref{eq:c23_n5}) cannot be satisfied.  
Thereafter, it increases and approaches $9/8$ at $k_2/k_1\to 0$, for $k_2/k_1>0$ monotonically decreases and  asymptotically approaches $25/24$ at $k_2/k_1\to \infty$.  
 Thus,  because the squared angular momentum does not have an upper bound but have the lower bound $j^2=0.841...$, we find that  it must run the range 
\begin{eqnarray}
j^2>0.841...\ . 
\end{eqnarray}
 
From this analysis, we can conclude  that  
the bi-axisymmetric and reflection-symmetric microstate geometry with five centers can have the  angular momentum of the range $0.841...<j^2<1$ as the BMPV black hole, while the microstate geometry with three centers cannot have.

\section{Summary and Discussions}\label{sec:summary}

In this paper, we have analyzed the solutions of the asymptotically flat, stationary, BPS microstate geometries with bi-axisymmetry in the five-dimensional minimal supergravity. 
Moreover, we have imposed additional reflection symmetry since this symmetry assumption extremely simplifies the expression of the solutions and enables us to solve the bubble equations. 
We  have also computed the conserved charges, the ADM mass, two ADM angular momenta, and $(n-1)$ magnetic fluxes through the bubbles between two centers. 
In particular, we have compared  the mass and angular momenta for the three-center solution and the five-center solution of microstate geometries with those of the BMPV black hole.
We have shown that the dimensionless angular momentum of the five-center microstate geometry does not have the upper bound but has the lower bound which is smaller than the angular momentum for the maximally spinning BMPV black hole, and hence there are the parameter region such that the microstate geometry has the same angular momentum as the BMPV black hole.

\medskip
In our present analysis, we have restricted ourselves to the reflection-symmetric microstate geometries for $n=3$ and $n=5$, but it is not trivial whether there exist the reflection-symmetric solutions with a larger number of centers ($n=7,9,\ldots$) which admit the same mass and angular momentum as the BMPV black hole or the microstate geometries for $n=3,5$. 
The bi-axisymmetric and reflection-symmetric microstate geometry with $n$ centers seems to have $(n+3)/2$ independent physical charges or fluxes [the mass $M$, the angular momentum $J_\psi$ or the $(n-1)/2$ magnetic fluxes $q[I_i]\ (i=1,\ldots,(n-1)/2)$], among which only $(n-1)/2$ are independent since the number of the parameters reduces to half due to reflection symmetry. 
The analysis for such microstate  geometries with $n\ge 7$ deserves future works.   
Moreover, it may be interesting to compare the five-center solution dealt with in this paper with the spherical  black holes having a topologically nontrivial domain of outer communication in Refs.~\cite{Kunduri:2014iga,Horowitz:2017fyg}, which can have  not only same asymptotic charges as the BMPV black hole but also different ones. 
The solution without the reflection symmetry should be compared with the supersymmetric black ring~\cite{Elvang:2004rt} and supersymmetric black lenses~\cite{Kunduri:2014kja,Breunholder:2017ubu,Tomizawa:2016kjh} which does not admit the limit to equal angular momenta. 
This may be an interest issue as our future study. 
Finally, we comment that the solutions of the five-dimensional minimal supergravity can be uplifted to the solutions of both type IIB supergravity and eleven supergravity~\cite{Buchel:2006gb,Gauntlett:2007ma}, and  
as discussed in Ref.~\cite{Gauntlett:2006ai}, such solutions are relevant for the most general four-dimensional  superconformal field theories (SCFTs) with holographic duals.  
This enables one to study some aspects of the dual strongly coupled thermal plasma with a non-zero R-charge chemical potential. 
Therefore, it might be physically interesting to study the fluid-dynamics of the thermal plasma of the SCFTs corresponding to the microstate geometries.




\acknowledgments
We thank Masaki Shigemori for useful comments. 
This work was supported by the Grant-in-Aid for Scientific Research (C) [JSPS KAKENHI Grant Number~17K05452] and the Grant-in-Aid for Scientific Research (C) [JSPS KAKENHI Grant Number~21K03560] and from the Japan Society for the Promotion of Science is also supported from Toyota Institute of Technology Fund for Research Promotion (A).




\end{document}